\documentclass[sigconf]{acmart}

\usepackage{amsmath}
\usepackage{booktabs}
\usepackage{graphicx}
\usepackage{enumitem}
\usepackage{subcaption}
\usepackage{placeins}
\usepackage{dblfloatfix}
\usepackage{tikz}
\usepackage{xcolor}
\usepackage{hyperref}

\usetikzlibrary{arrows.meta}

\graphicspath{{./}}
\title{TARE: Tail Aware Evaluation of HPC Job Runtime Prediction}

\author{Haili Xiao}
\authornote{Both authors contributed equally to this work as co-first authors.}
\affiliation{
  \institution{CNIC, CAS}
  \country{China}
}
\email{haili@sccas.cn}

\author{Can Wu}
\authornotemark[1]
\affiliation{
  \institution{CNIC, CAS}
  \country{China}
}
\email{wucan@sccas.cn}

\author{Shasha Lu}
\affiliation{
  \institution{CNIC, CAS}
  \country{China}
}
\email{lusha721@sccas.cn}

\author{Xiaoning Wang}
\affiliation{
  \institution{CNIC, CAS}
  \country{China}
}
\email{wxn@sccas.cn}

\author{Yining Zhao}
\affiliation{
  \institution{CNIC, CAS}
  \country{China}
}
\email{zhaoyn@sccas.cn}

\author{Rong He}
\affiliation{
  \institution{CNIC, CAS}
  \country{China}
}
\email{herong@sccas.cn}

\acmConference[ICPP '26]{The 55th International Conference on Parallel Processing}{September 28--October 1, 2026}{Singapore}
\acmBooktitle{The 55th International Conference on Parallel Processing (ICPP '26), September 28--October 1, 2026, Singapore}

\begin{document}

\begin{abstract}
Runtime estimates affect reservation quality, backfilling opportunities, and queue delay in HPC schedulers. Under heavy tailed workloads, however, averaging over jobs can misrepresent scheduling impact because a small fraction of jobs dominates resource usage. This paper presents an empirical evaluation methodology for HPC job runtime prediction that focuses on the tail, combining GeoAccuracy weighted by resource usage with decile and split analyses. Using production traces from NREL Eagle and ALCF Mira/Intrepid, we compare \texttt{XGBoost} and \texttt{Last2} against the user provided walltime estimate at submission (\texttt{UserReq}). Across all three datasets, evaluation focused on the tail changes the offline conclusion: MeanAccuracy keeps the methods relatively close, whereas GeoAccuracy reveals clearer separation and makes \texttt{UserReq}'s strength in the upper tail visible. In the top decile, \texttt{UserReq} achieves the highest GeoAccuracy and lowest underestimation rate on all three datasets, and this pattern remains stable across rolling splits. We then translate this signal into a simple hybrid scheduling policy that keeps \texttt{XGBoost} for most jobs and routes the top decile by \texttt{proxy\_cost} at submission to \texttt{UserReq}. Online replay on four production queues reduces mean wait time by up to 8\% and increases backfilled jobs by 50\%--115\%. These results show that offline evaluation focused on the tail better characterizes prediction quality relevant to scheduling and informs scheduling policy design.

\end{abstract}



\maketitle

\section{Introduction}

In batch scheduled HPC systems, users submit jobs with requested resources such as node counts and walltime limits, and those jobs wait in queue until the scheduler selects them for execution. Under \texttt{FCFS} with \texttt{EASY} backfilling, the scheduler uses a runtime estimate to reserve resources for the job at the head of the queue and to determine whether later jobs can start early without delaying that reservation~\cite{lifka1995ANLIBM, feitelson1998UtilizationPredictability}. Runtime estimation therefore affects reservation quality, backfilling opportunities, and queue performance directly. Overestimation can leave schedulable gaps unused, while underestimation can make planned reservations less reliable. For this reason, HPC job runtime prediction has long been studied as a scheduling problem rather than only as a forecasting problem~\cite{smith1999UsingRunTime, tsafrir2007BackfillingUsing}.

\begin{figure}[!htb]
\centering
\resizebox{0.98\columnwidth}{!}{\begin{tikzpicture}[font=\scriptsize, line cap=round, line join=round]
\draw[rounded corners=2pt, draw=black!45, fill=black!2, line width=0.45pt] (0,0) rectangle (6.55,2.75);
\node[font=\bfseries\scriptsize, align=center] at (3.275,2.32) {GeoAccuracy reveals clearer differences};
\foreach \x/\h/\shade in {1.02/0.68/68,1.38/0.55/44,1.74/0.36/82}{
  \draw[draw=blue!45!black, fill=blue!\shade!black!8, line width=0.26pt] (\x,0.52) rectangle ++(0.24,\h);
}
\draw[-{Stealth[length=4pt,width=5pt]}, draw=black!55, line width=0.6pt] (2.79,0.88) -- (3.47,0.88);
\foreach \x/\h/\shade in {4.28/0.66/66,4.64/0.30/40,5.00/1.34/84}{
  \draw[draw=green!45!black, fill=green!\shade!black!8, line width=0.26pt] (\x,0.52) rectangle ++(0.24,\h);
}
\node[font=\tiny] at (1.58,0.26) {MeanAccuracy};
\node[font=\tiny] at (4.84,0.26) {GeoAccuracy};
\end{tikzpicture}}
\caption{Offline evaluation focused on the tail gives new implications for scheduling policy design (see Figure~\ref{fig:hybrid-policy-flow}).}
\Description{A compact conceptual diagram. The box title says that GeoAccuracy reveals clearer differences. Inside, a small bar group labeled MeanAccuracy shows only modest separation, a right-pointing arrow indicates the comparison shift, and a taller bar group labeled GeoAccuracy shows clearer separation.}
\label{fig:tare-concept}
\end{figure}
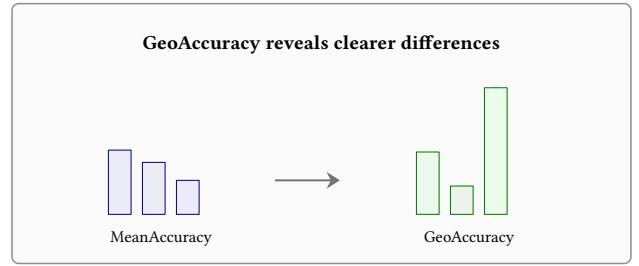

Because runtime estimates directly affect scheduling decisions, evaluating their quality is itself a scheduling problem. Most runtime prediction studies focus on predictor design and report aggregate accuracy at the job level. But the practical significance of a prediction error depends on which jobs incur it. If a workload has heavy tails, a small fraction of jobs can dominate resource usage. That property is central here: averaging over jobs can obscure the errors with the greatest scheduling consequences. We therefore ask whether weighting accuracy by resource usage reveals predictor differences that job averaged evaluation keeps much closer together. Figure~\ref{fig:tare-concept} previews this comparison. We compare the weighted geometric metric GeoAccuracy, $A_{\mathrm{geo}}$, with its arithmetic mean counterpart over jobs, MeanAccuracy, $A_{\mathrm{mean}}$. We study two runtime predictors and compare them with the user provided walltime estimate at job submission:
\begin{itemize}[leftmargin=1.5em]
\item \texttt{XGBoost}, a machine learning predictor;
\item \texttt{Last2}, a heuristic baseline; and
\item \texttt{UserReq}, the user provided walltime estimate at job submission.
\end{itemize}
We compare these three methods on three production workload datasets: \texttt{Eagle} from NREL and \texttt{Mira}/\texttt{Intrepid} from the Argonne Leadership Computing Facility (ALCF)~\cite{menear2023MasteringHPC, patel2020JobCharacteristics}. All three come from large production HPC systems at national laboratories. Shared preprocessing yields aligned prediction sources. We then apply a consistent offline evaluation across all three datasets and use its findings to design an online replay study on \texttt{Mira} and \texttt{Intrepid}.

\begin{figure}[t]
\centering
\begin{subfigure}[t]{\columnwidth}
    \centering
    \includegraphics[width=\textwidth]{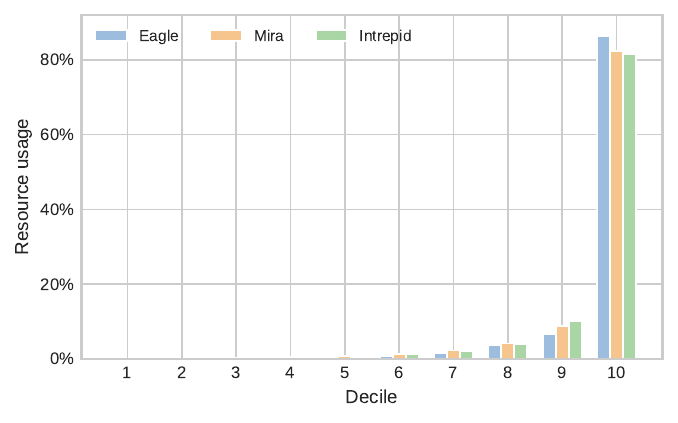}
    \caption{Equal sized decile.}
    \label{fig:tail-structure-a}
\end{subfigure}
\par\smallskip
\begin{subfigure}[t]{\columnwidth}
    \centering
    \includegraphics[width=\textwidth]{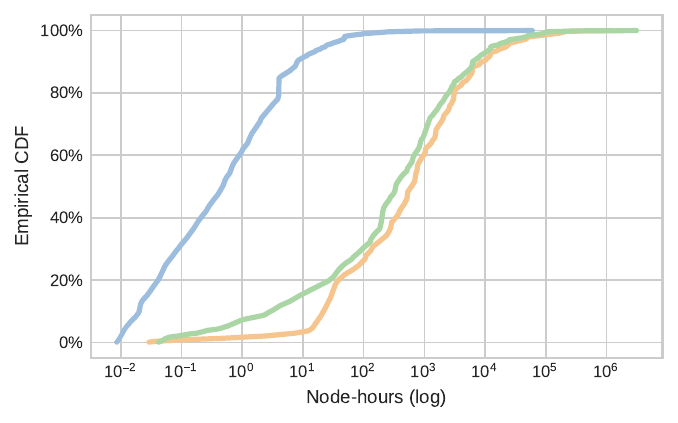}
    \caption{Empirical CDF of node hours.}
    \label{fig:tail-structure-b}
\end{subfigure}
\caption{Evidence of a heavy tail. The highest 10\% alone contributes about 81\%--87\% of total resource usage across the three datasets.}
\Description{Two stacked single-column plots. The first is a grouped bar chart over equally sized deciles, with separate bars for Eagle, Mira, and Intrepid; the y-axis is resource usage share, showing strong concentration in the highest decile across all three datasets. The second is a log-scale empirical CDF plot of node-hours, showing the same concentration from a continuous-distribution perspective.}
\label{fig:tail-structure}
\end{figure}

Before comparing metrics, we verify the workload property that motivates the study in these three traces. A heavy tailed resource usage pattern is not a new discovery of this paper. Prior HPC workload characterization studies have repeatedly reported strong skew and heterogeneity in job size, resource usage, and workload structure on large HPC systems~\cite{patel2020JobCharacteristics, wang2021UserlevelWorkload, li2023AnalyzingResource}. Our purpose is to show that the same concentration appears in \texttt{Eagle}, \texttt{Mira}, and \texttt{Intrepid}, and to quantify its magnitude. Figure~\ref{fig:tail-structure} shows that the highest decile alone contributes about 81\%--87\% of total resource usage across the three datasets. Because HPC resources are scarce and scheduler impact is dominated by large, long running jobs, that concentration makes upper-tail behavior especially important. The CDF view in Figure~\ref{fig:tail-structure-b} shows the same concentration from a continuous distribution perspective: \texttt{Eagle} sits farther to the upper left because its jobs consume much smaller node hours, whereas \texttt{Mira} and \texttt{Intrepid} lie farther to the lower right and remain close because the two ALCF traces have similarly large job scales.

Under a workload with heavy tails, averages over jobs give every job the same influence, whereas averages weighted by resource usage give more influence to the jobs that consume most node time. As a result, the same method comparison can support different overall conclusions. In our data, $A_{\mathrm{mean}}$ keeps the three methods much closer together, whereas $A_{\mathrm{geo}}$ reveals much clearer differences and makes \texttt{UserReq}'s strength in the upper tail visible on all three datasets. This result is notable because \texttt{UserReq} is the user provided walltime estimate at job submission rather than a learned predictor, but the explanation is that $A_{\mathrm{geo}}$ emphasizes the upper tail of resource usage, where \texttt{UserReq} performs better. This difference also remains stable across most rolling split days, so it is not an artifact of a small sample.

The paper follows a three part experimental workflow: shared data preparation, offline evaluation, and online simulation. The offline evaluation weights errors by resource usage and analyzes the upper tail explicitly; the online simulation then tests whether this offline upper-tail signal translates into scheduling outcomes. This emphasis comes from the workload weights and upper-tail analysis, not from any special outlier sensitivity of the geometric mean alone.

The contribution is empirical and evaluative rather than algorithmic. We do not propose a new predictor. Instead, this study makes three contributions:
\begin{enumerate}[leftmargin=1.5em]
\item We present an empirical evaluation methodology for HPC job runtime prediction that focuses on the tail. It combines GeoAccuracy weighted by resource usage with decile and split level analyses, so offline evaluation reflects the jobs that dominate scheduling impact rather than averages over jobs alone.
\item Using three production workload datasets, we show that evaluation focused on the tail changes the overall comparison of runtime predictors. MeanAccuracy keeps the methods much closer together, whereas GeoAccuracy reveals clearer separation and makes \texttt{UserReq}'s strength in the resource dominant upper tail visible. In the top decile, \texttt{UserReq} achieves the highest GeoAccuracy and the lowest underestimation rate on all three datasets, and this advantage remains stable across rolling split days.
\item We translate this offline signal into a simple hybrid scheduling policy and show in online replay that it improves scheduling performance. The policy keeps \texttt{XGBoost} for most jobs and routes the top decile of jobs by \texttt{proxy\_cost}, computed from requested nodes and walltime at submission, to \texttt{UserReq}. Across the short and long production queues on two HPC systems, this policy reduces mean wait time by up to 8\% and increases backfilled jobs by 50\%--115\%.
\end{enumerate}

The rest of the paper is organized as follows. Section~2 presents the methodology, including the offline evaluation metrics and the offline analysis procedure. Section~3 describes the experimental setup, including data preparation, datasets, predictors, and the rolling split offline evaluation procedure. Section~4 presents the offline results, including the overall metric comparison, analysis of the upper tail, and robustness at the split level. Section~5 presents the online simulation, including its design and results. Section~6 reviews related work, and Section~7 concludes the paper.

\section{Methodology}

\begin{figure*}[!t]
\centering
\resizebox{\textwidth}{!}{\begin{tikzpicture}[font=\scriptsize, >=Latex, line cap=round, line join=round]

\tikzset{
  stage/.style={rounded corners=3pt, fill=black!2, draw=black!35, line width=0.45pt},
  stage_title/.style={rounded corners=2pt, fill=black!10, draw=black!35, line width=0.4pt, font=\bfseries\scriptsize, inner sep=3pt},
  data_box/.style={rounded corners=2pt, fill=blue!5, draw=blue!45!black, line width=0.4pt, align=center, inner sep=4pt},
  offline_box/.style={rounded corners=2pt, fill=green!6, draw=green!45!black, line width=0.4pt, align=center, inner sep=4pt},
  online_box/.style={rounded corners=2pt, fill=orange!8, draw=orange!55!black, line width=0.4pt, align=center, inner sep=4pt},
  flow/.style={->, line width=0.7pt, draw=black!65},
  linklabel/.style={font=\tiny, fill=white, inner sep=1pt}
}

\draw[stage] (0.00,0.00) rectangle (4.85,5.05);
\draw[stage] (5.35,0.00) rectangle (10.20,5.05);
\draw[stage] (10.70,0.00) rectangle (15.55,5.05);

\node[stage_title] at (2.425,4.72) {Data Preparation};
\node[stage_title] at (7.775,4.72) {Offline Evaluation};
\node[stage_title] at (13.125,4.72) {Online Simulation};

\node[data_box, minimum width=3.95cm, minimum height=0.82cm] (raw) at (2.425,3.80)
  {Raw job traces\\\texttt{Eagle}, \texttt{Mira}, \texttt{Intrepid}};
\node[data_box, minimum width=3.95cm, minimum height=1.05cm] (prep) at (2.425,2.42)
  {Shared preprocessing\\keep valid terminal states, required attributes,\\\texttt{runtime > 30 s}};
\node[data_box, minimum width=3.95cm, minimum height=0.92cm] (pred) at (2.425,1.00)
  {Aligned prediction sources\\\texttt{XGBoost}, \texttt{Last2}, \texttt{UserReq}};

\draw[flow] (raw.south) -- (prep.north);
\draw[flow] (prep.south) -- (pred.north);

\node[offline_box, minimum width=4.00cm, minimum height=0.90cm] (split) at (7.775,3.80)
  {Daily rolling splits\\100-day train, 1-day test};
\node[offline_box, minimum width=4.00cm, minimum height=1.05cm] (analysis) at (7.775,2.38)
  {Offline analyses\\overall metrics, split robustness,\\equally sized deciles};
\node[offline_box, minimum width=4.00cm, minimum height=1.05cm] (tail) at (7.775,0.95)
  {Offline takeaway\\upper-tail jobs dominate resource usage;\\\texttt{UserReq} becomes stronger in the top decile};

\draw[flow] (split.south) -- (analysis.north);
\draw[flow] (analysis.south) -- (tail.north);

\node[online_box, minimum width=4.00cm, minimum height=1.00cm] (replay) at (13.125,3.78)
  {Queue-year trace replay\\\texttt{Mira}/\texttt{Intrepid} short and long production queues\\\texttt{FCFS} + \texttt{EASY} backfilling};
\node[online_box, minimum width=4.00cm, minimum height=1.00cm] (policy) at (13.125,2.26)
  {Hybrid-p90 policy\\top 10\% by \texttt{proxy\_cost} $\rightarrow$ \texttt{UserReq},\\otherwise \texttt{XGBoost}};
\node[online_box, minimum width=4.00cm, minimum height=0.90cm] (outcomes) at (13.125,0.88)
  {Reported outcomes\\mean wait time, backfilled-job count};

\draw[flow] (replay.south) -- (policy.north);
\draw[flow] (policy.south) -- (outcomes.north);

\draw[flow] (4.95,2.52) -- (5.25,2.52);

\draw[flow] (10.30,2.52) -- (10.60,2.52);

\end{tikzpicture}}
\caption{Overview of the experimental workflow. (1) Data preparation produces aligned prediction sources for three traces. (2) Offline evaluation uses daily rolling splits; its finding on the upper tail then informs the hybrid policy used in online simulation. (3) Online simulation replays four production queues from \texttt{Mira} and \texttt{Intrepid} and reports mean wait time and the count of backfilled jobs.}
\Description{A three-stage workflow diagram. The first stage, Data Preparation, starts from raw job traces for Eagle, Mira, and Intrepid, applies shared preprocessing with valid terminal states, required attributes, and runtime greater than 30 seconds, and produces aligned prediction sources XGBoost, Last2, and UserReq. The second stage, Offline Evaluation, applies daily rolling splits with a 100-day training window and a 1-day test window, then performs overall-metric, split-robustness, and decile analyses, and concludes that upper-tail jobs dominate resource usage and that UserReq becomes stronger in the top decile. The third stage, Online Simulation, uses queue-year trace replay on the short and long production queues of Mira and Intrepid under FCFS with EASY backfilling, applies a Hybrid-p90 policy that routes the top decile by proxy_cost to UserReq and the rest to XGBoost, and reports mean wait time and backfilled-job count. Arrows connect the three stages from left to right.}
\label{fig:experiment-overview}
\end{figure*}
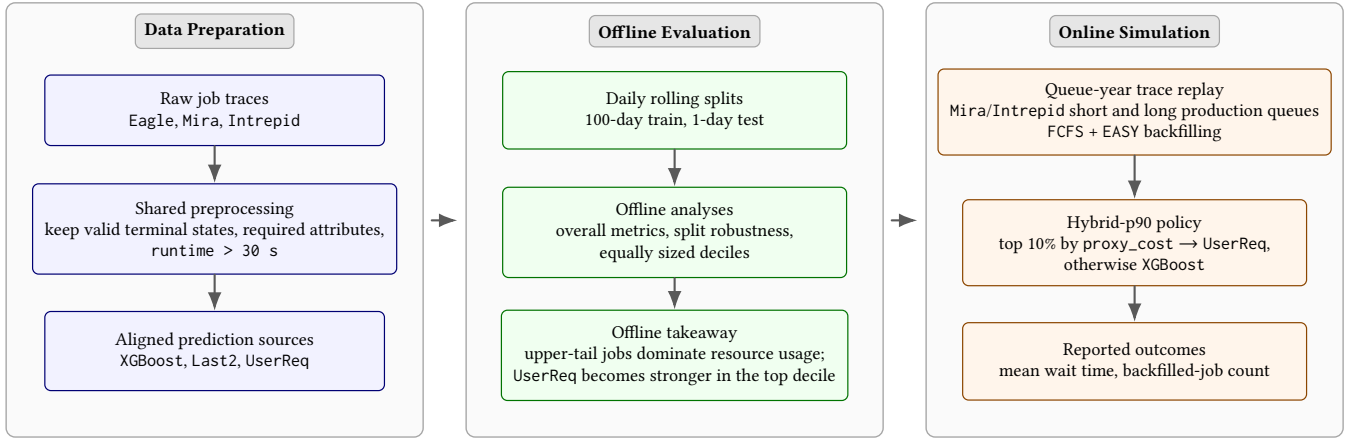

\subsection{Offline Evaluation Metrics}

In the batch queue setting, a submitted job carries requested resources and a user walltime estimate, then later yields an observed actual runtime after execution. For job $i$, let $y_i$ denote the actual runtime, $\hat{y}_i$ the predicted runtime, and $c_i = \texttt{nodes\_req} \times \texttt{runtime\_act}$ denote the resource usage used as the offline weight.

We use multiplicative error rather than additive error because HPC job runtimes span a wide range of scales. An additive error measured in seconds is therefore not directly comparable across jobs: the same absolute miss can be minor for a long job but severe for a short job. Taking a log ratio turns multiplicative misses into additive quantities, so ratio errors can be averaged cleanly across jobs. It also treats paired ratio errors symmetrically, so predicting $2y_i$ and $y_i/2$ yields the same penalty. This keeps relative prediction quality separate from the weights by resource usage introduced later in $A_{\mathrm{geo}}$. The basic log ratio error is
\begin{equation}
e_i = \left| \log \frac{\hat{y}_i}{y_i} \right|.
\end{equation}
Because the shared preprocessing keeps only jobs with observed runtime above 30 s and the evaluation retains only valid positive predictions, every analyzed row satisfies $y_i > 0$ and $\hat{y}_i > 0$. We therefore do not need a separate branch for zero runtime in the metric definition.

The accuracy score for each job is
\begin{equation}
a_i = \exp(-e_i) = \min\left(\frac{\hat{y}_i}{y_i}, \frac{y_i}{\hat{y}_i}\right).
\end{equation}
This yields the familiar arithmetic mean metric over jobs
\begin{equation}
A_{\mathrm{mean}} = \frac{1}{N} \sum_{i=1}^{N} a_i.
\end{equation}
Throughout the paper, we refer to $A_{\mathrm{mean}}$ as MeanAccuracy.

To emphasize jobs with high resource usage, we also define the log error weighted by resource usage
\begin{equation}
L_{\mathrm{usage}} = \frac{\sum_i c_i e_i}{\sum_i c_i}.
\end{equation}
The corresponding weighted geometric metric is
\begin{equation}
A_{\mathrm{geo}} = \exp(-L_{\mathrm{usage}})
= \prod_i a_i^{\,c_i / \sum_j c_j},
\end{equation}
which is the weighted geometric mean of the job level accuracy scores $a_i$. Throughout the paper, we refer to $A_{\mathrm{geo}}$ as GeoAccuracy.

The log form is what makes the geometric mean natural here. Averaging $e_i$ summarizes a typical multiplicative miss on an additive scale, and exponentiating back returns the result to the same $(0,1]$ accuracy scale as $A_{\mathrm{mean}}$. GeoAccuracy is therefore not an arbitrary alternative summary: it is the consistent aggregate of multiplicative error under weights by resource usage.

The emphasis on the workload tail comes from the weights $c_i$, not from the use of a geometric mean by itself. GeoAccuracy emphasizes the tail because it weights jobs by resource usage and places most mass on the upper tail of that distribution. It does not make the metric inherently more sensitive to extreme error outliers.

This distinction matters because MeanAccuracy and GeoAccuracy answer different questions. MeanAccuracy is an average over jobs. GeoAccuracy is weighted by resource usage and therefore emphasizes the part of the workload whose resource usage most strongly affects scheduling outcomes.

To explain the overall metric difference, we also evaluate predictors on equally sized deciles of offline resource usage. Within each dataset, we sort jobs by $c_i$ and divide them into ten equally sized groups; the tenth group is the top decile. This decile view complements the overall metrics: it explains why GeoAccuracy can reveal clearer differences than MeanAccuracy and localizes the upper tail discussed later in the paper. When the online section later refers to a p90 or top decile routing rule, it uses the same percentile language but applies it to the proxy $\texttt{proxy\_cost}$ at submission within each queue-year partition, rather than to the offline resource usage variable $c_i$.

We also report the underestimation rate
\begin{equation}
\mathrm{UnderRate} = \frac{1}{N} \sum_i \mathbf{1}[\hat{y}_i < y_i].
\end{equation}

UnderRate is useful because it exposes a different failure mode from the accuracy metrics. A predictor can achieve competitive MeanAccuracy or GeoAccuracy while still underestimating actual runtime too often. We therefore treat MeanAccuracy, GeoAccuracy, and UnderRate as complementary rather than redundant summaries.

Accordingly, the offline evaluation proceeds in three steps. It first compares the overall metrics on the valid rows of each model, then analyzes equally sized deciles of resource usage within each dataset to isolate the highest decile, and finally checks whether the corresponding metric gaps at the split level remain stable across rolling split days. These offline results then motivate the online simulation reported later.

\section{Experimental Setup}

Figure~\ref{fig:experiment-overview} summarizes the experimental workflow. This section covers its first two parts: shared data preparation and rolling split offline evaluation on all three datasets. Section~5 then uses the same traces in online replay, which preserves the observed chronology of job arrivals and changes only the planned runtimes seen by the scheduler.

\subsection{Data Preparation and Date Ranges}\label{sec:data-preparation}

We study three production traces from systems at national laboratories: \texttt{Eagle} at NREL and \texttt{Mira}/\texttt{Intrepid} at ALCF~\cite{menear2023MasteringHPC, patel2020JobCharacteristics}. Shared data preparation organizes the raw traces around common submit, start, end, and resource request fields; removes invalid or incomplete rows while retaining jobs with valid terminal states; constructs the inputs available at submission that are used by the predictors, including each user's recent job history needed by \texttt{Last2}; applies the common runtime filter \texttt{runtime > 30 s}; aligns the three prediction sources on consistent jobs; and defines rolling split days after a 100 day warm up window.

Their raw submit time spans are 2018-11 to 2023-02 for \texttt{Eagle}, 2014-10 to 2018-12 for \texttt{Mira}, and 2009-12 to 2013-12 for \texttt{Intrepid}. Following the prior Eagle runtime prediction workflow~\cite{menear2023MasteringHPC}, the offline evaluation uses daily rolling splits with a 100 day training window, so evaluated split days begin only after the first 100 days of history. The resulting ranges of split days are 2019-02 to 2023-02 for \texttt{Eagle}, 2015-02 to 2018-12 for \texttt{Mira}, and 2010-03 to 2013-12 for \texttt{Intrepid}. After filtering and alignment of prediction sources, the offline evaluation contains 7,267,468 jobs and 1,435 rolling splits for \texttt{Eagle}, 242,753 jobs and 1,425 splits for \texttt{Mira}, and 271,713 jobs and 1,379 splits for \texttt{Intrepid}.

\subsection{Runtime Predictors and User Estimate}

The two runtime predictors and the user provided walltime estimate at job submission play distinct roles. \texttt{XGBoost} is a gradient boosted tree predictor and a strong representative choice for tabular runtime prediction tasks~\cite{chen2016XGBoost, chen2020RuntimePrediction, menear2023MasteringHPC}. \texttt{Last2} is a heuristic baseline that predicts a job as the mean runtime of the same user's two most recent jobs in the training window, with fallback to the global recent job mean when that user has no history~\cite{tsafrir2007BackfillingUsing}. \texttt{UserReq} uses the submitted walltime request \texttt{wallclock\_req} directly.

\subsection{Offline Evaluation Procedure}

Each rolling split trains on jobs whose \texttt{end\_time} lies in the previous 100 days and tests on jobs whose \texttt{submit\_time} lies in the next 1 day. For split day $t$, the training window is $[t-100\text{ days}, t)$ in \texttt{end\_time} and the test window is $[t, t+1\text{ day}]$ in \texttt{submit\_time}. Advancing $t$ by one day creates a time ordered sequence of metrics rather than a single random holdout. Decile boundaries are computed once per dataset by sorting jobs by resource usage and splitting them into ten equally sized groups; the same boundaries are then applied to all models.

At the split level, the analysis unit is each rolling split day rather than individual jobs, and comparisons use the split days where both predictors have valid offline metrics. For each metric, we summarize the fraction of common split days where one predictor outperforms the other: higher values win for metrics such as $A_{\mathrm{mean}}$ and $A_{\mathrm{geo}}$, while lower values win for metrics such as UnderRate.

\section{Offline Results}

\subsection{Evaluation Focused on the Tail Changes the Offline Conclusion}

\begin{figure*}[t]
\centering
\includegraphics[width=\textwidth]{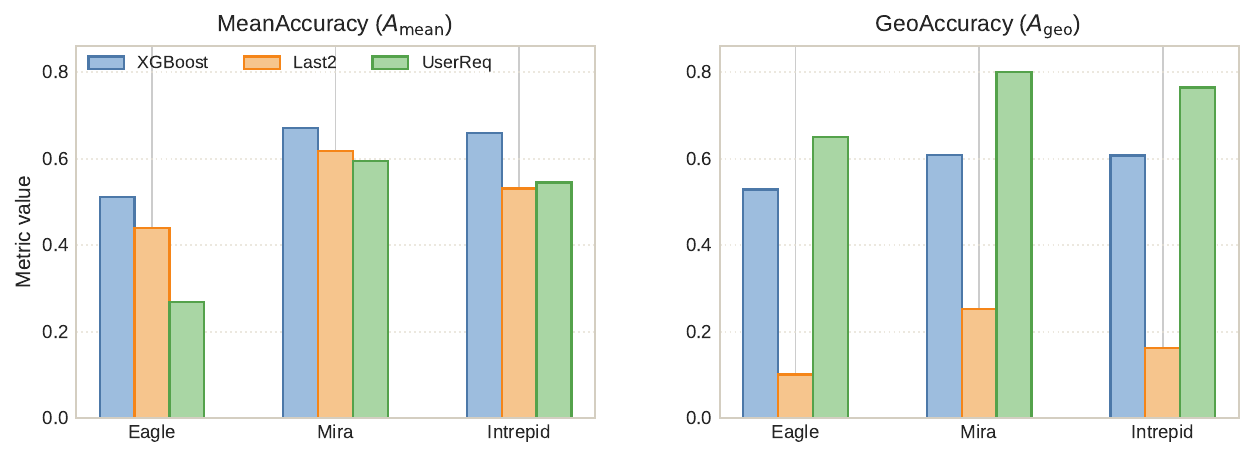}
\captionsetup{justification=centering,singlelinecheck=true}
\caption{Overall offline metrics. GeoAccuracy reveals much clearer differences than MeanAccuracy does.}
\Description{A two-panel grouped bar chart comparing MeanAccuracy and GeoAccuracy. In both panels, bars are grouped by dataset and colored by model. The MeanAccuracy panel keeps the bars relatively close together, while the GeoAccuracy panel shows substantially clearer differentiation among XGBoost, Last2, and UserReq.}
\label{fig:overall-metrics}
\end{figure*}

With the heavy tail workload property established in Section~1, Table~\ref{tab:overall-metrics} and Figure~\ref{fig:overall-metrics} report the main overall metrics. The key result is that evaluation focused on the tail changes the offline conclusion. Under $A_{\mathrm{mean}}$, the methods remain much closer together, whereas $A_{\mathrm{geo}}$ separates them much more clearly. The gap between the maximum and minimum across the three models increases from 0.242 to 0.549 in \texttt{Eagle}, from 0.077 to 0.549 in \texttt{Mira}, and from 0.129 to 0.603 in \texttt{Intrepid} when moving from $A_{\mathrm{mean}}$ to $A_{\mathrm{geo}}$.

This larger separation makes the effect of the workload tail visible at the overall metric level. Under $A_{\mathrm{mean}}$, the three methods stay close together, especially in \texttt{Mira} and \texttt{Intrepid}. Under $A_{\mathrm{geo}}$, \texttt{Last2} drops sharply and \texttt{UserReq} pulls clearly away from the others. The ranking change is therefore a consequence of stronger separation in the tail; the more fundamental point is that evaluation weighted by resource usage reveals much larger performance differences among the methods from a scheduling perspective.

\begin{table}[!b]
\captionsetup{justification=centering,singlelinecheck=true}
\caption{Overall offline metrics.}
\label{tab:overall-metrics}
\centering
\begingroup
\setlength{\tabcolsep}{2.5pt}
\renewcommand{\arraystretch}{0.92}
\scriptsize
\resizebox{0.82\columnwidth}{!}{\begin{tabular}{llrrr}
\toprule
Dataset & Model & $A_{\mathrm{mean}}$ & $A_{\mathrm{geo}}$ & $\mathrm{UnderRate}$ \\
\midrule
Eagle & XGBoost & 0.511 & 0.529 & 39.9\% \\
 & Last2 & 0.440 & 0.101 & 52.5\% \\
 & UserReq & 0.269 & 0.650 & 11.0\% \\
Mira & XGBoost & 0.671 & 0.609 & 54.0\% \\
 & Last2 & 0.617 & 0.252 & 50.3\% \\
 & UserReq & 0.594 & 0.801 & 17.1\% \\
Intrepid & XGBoost & 0.660 & 0.608 & 48.7\% \\
 & Last2 & 0.531 & 0.162 & 55.8\% \\
 & UserReq & 0.545 & 0.765 & 11.5\% \\
\bottomrule
\end{tabular}
}
\endgroup
\end{table}

\subsection{Analysis of the Upper Tail Explains Why the Conclusion Changes}

\begin{figure*}[!b]
\centering
\includegraphics[width=\textwidth]{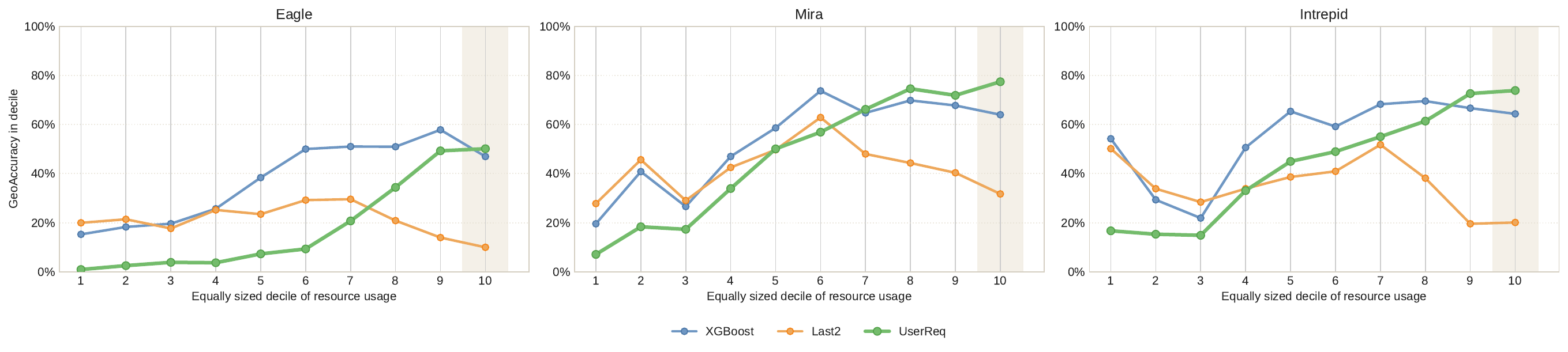}
\caption{Why and where \texttt{UserReq} becomes a useful estimate. \texttt{UserReq} is weak on many jobs with low resource usage but improves sharply toward the higher deciles, eventually becoming competitive with or stronger than \texttt{XGBoost} and \texttt{Last2} in the upper tail.}
\Description{Three side-by-side line charts for Eagle, Mira, and Intrepid showing GeoAccuracy by equally sized deciles of resource usage for XGBoost, Last2, and UserReq. The UserReq line starts low in early deciles and rises strongly toward later deciles, especially the final decile.}
\label{fig:userreq-decile-trajectory}
\end{figure*}

Figure~\ref{fig:userreq-decile-trajectory} shows the decile pattern behind the changed offline conclusion. \texttt{UserReq} is not uniformly strong across the workload. In the lowest deciles of resource usage, it is often the weakest of the three methods because many small jobs are requested conservatively and therefore look inaccurate under a multiplicative metric. This pattern changes sharply as resource usage increases.

From the first to the tenth decile of resource usage, \texttt{UserReq} GeoAccuracy rises from 0.009 to 0.501 in \texttt{Eagle}, from 0.071 to 0.774 in \texttt{Mira}, and from 0.167 to 0.738 in \texttt{Intrepid}. By the upper tail, \texttt{UserReq} closes the gap with \texttt{XGBoost} and becomes stronger in the region with the highest resource usage: the tenth decile in \texttt{Eagle}, the seventh through tenth deciles in \texttt{Mira}, and the ninth and tenth deciles in \texttt{Intrepid}.

This decile shift explains why GeoAccuracy and MeanAccuracy support different offline conclusions. MeanAccuracy gives equal weight to the many jobs with low resource usage where \texttt{UserReq} is weak. GeoAccuracy gives much more weight to the upper deciles, where \texttt{UserReq} becomes progressively more accurate and where resource usage is concentrated. The overall metric difference is therefore driven by a clear change in predictor behavior across deciles.

\subsection{Top Decile Results Localize the Advantage Relevant to Scheduling}

Figure~\ref{fig:userreq-decile-trajectory} already shows the decile trend. The highest decile localizes the advantage relevant to scheduling because it is the part of the workload that carries most of the resource usage.

\begin{table*}[t]
\caption{Top decile offline metrics. Jobs are split into ten equally sized deciles by offline resource usage; ``Top decile usage'' reports the share of total usage carried by the highest decile in each dataset.}
\label{tab:top-decile-metrics}
\centering
\begingroup
\setlength{\tabcolsep}{7pt}
\small
\begin{tabular}{llrrr}
\toprule
Dataset & Model & \shortstack[r]{Top-decile\\usage} & \shortstack[r]{Top-decile\\GeoAccuracy} & \shortstack[r]{Top-decile\\$\mathrm{UnderRate}$} \\
\midrule
Eagle & XGBoost & 86.5\% & 0.470 & 72.7\% \\
 & Last2 &  & 0.100 & 82.5\% \\
 & UserReq &  & 0.501 & 34.1\% \\
Mira & XGBoost & 82.3\% & 0.640 & 77.9\% \\
 & Last2 &  & 0.317 & 74.0\% \\
 & UserReq &  & 0.774 & 32.2\% \\
Intrepid & XGBoost & 81.4\% & 0.643 & 75.7\% \\
 & Last2 &  & 0.201 & 80.3\% \\
 & UserReq &  & 0.738 & 26.0\% \\
\bottomrule
\end{tabular}

\endgroup
\end{table*}

Table~\ref{tab:top-decile-metrics} reports the exact metrics for the top decile. In every dataset, \texttt{UserReq} achieves the highest GeoAccuracy in the top decile and the lowest underestimation rate. Relative to \texttt{XGBoost}, its GeoAccuracy in the top decile rises from 0.470 to 0.501 in \texttt{Eagle}, from 0.640 to 0.774 in \texttt{Mira}, and from 0.643 to 0.738 in \texttt{Intrepid}. Over the same datasets, UnderRate falls from 72.7\% to 34.1\%, from 77.9\% to 32.2\%, and from 75.7\% to 26.0\%, which corresponds to reductions of 38.6, 45.7, and 49.7 percentage points, respectively. It is also the only method that performs best on both GeoAccuracy and UnderRate in the top decile on all three datasets.

Across datasets, \texttt{XGBoost} remains competitive on GeoAccuracy in the tail, but its underestimation rate stays much higher; \texttt{Last2} is weaker on both tail metrics. This pattern across datasets is already clear from the reported values.

Taken together, these top decile results show a consistent pattern in the upper tail with clear implications for scheduling. The same prediction source that looks weak on many small jobs becomes both more accurate and less prone to underestimation on the jobs that dominate resource usage. That is the direct offline signal behind the hybrid policy studied later: if one prediction source is simultaneously safer and more accurate in the upper tail that matters most for scheduling, then routing that tail differently becomes a natural online design choice.

\subsection{Results at the Split Level Show That the Signal Is Stable}

Following the protocol at the split level in Section~3, we treat each rolling split day as one analysis unit and compare \texttt{XGBoost} with \texttt{UserReq} only on the split days where both predictors have valid offline metrics. Table~\ref{tab:split-robustness} reports the resulting win rates across split days. The signal at the split level remains stable under both metrics, but its direction depends on the metric: $A_{\mathrm{mean}}$ usually favors \texttt{XGBoost}, whereas $A_{\mathrm{geo}}$ usually favors \texttt{UserReq}. \texttt{XGBoost} has the higher $A_{\mathrm{mean}}$ on 67.3\%--87.4\% of split days, while \texttt{UserReq} has the higher $A_{\mathrm{geo}}$ on 68.9\%--81.0\% of split days. The reversal at the split level is therefore not coming from a few isolated days; it is the dominant direction across the rolling sequence.

\begin{table}[H]
\captionsetup{justification=centering,singlelinecheck=true}
\caption{Robustness at the split level, \texttt{XGBoost} vs.\ \texttt{UserReq}.}
\label{tab:split-robustness}
\centering
\begingroup
\renewcommand{\arraystretch}{1.02}
\small
\begin{tabular*}{0.96\columnwidth}{@{\extracolsep{\fill}}lrrr}
\toprule
Dataset & \shortstack[r]{XGBoost higher\\$A_{\mathrm{mean}}$} & \shortstack[r]{UserReq higher\\$A_{\mathrm{geo}}$} & \shortstack[r]{UserReq lower\\$\mathrm{UnderRate}$} \\
\midrule
Eagle & 87.3\% & 68.9\% & 97.4\% \\
Mira & 67.3\% & 81.0\% & 98.9\% \\
Intrepid & 87.4\% & 79.1\% & 99.6\% \\
\bottomrule
\end{tabular*}

\endgroup
\end{table}

\texttt{UserReq}'s advantage on underestimation is even more stable. It achieves the lower UnderRate in 97.4\% of split days on \texttt{Eagle}, 98.9\% on \texttt{Mira}, and 99.6\% on \texttt{Intrepid}. Variation at the split level is still expected because each day has a different test set size and a different upper tail composition. Even so, the dominant direction is consistent across the rolling sequence. The clearer differentiation under GeoAccuracy, together with \texttt{UserReq}'s much lower underestimation rate, is therefore a stable effect at the workload level rather than an artifact of a few extreme days.

\FloatBarrier

\section{Online Simulation}

The offline results show that \texttt{UserReq} becomes stronger precisely in the part of the workload that dominates resource usage and therefore matters most for scheduling. This section turns that finding into an online test: Section~5.1 defines the hybrid scheduling policy, Section~5.2 describes the replay setup, and Section~5.3 reports the resulting behavior at the queue level.

\subsection{Hybrid Scheduling Policy}

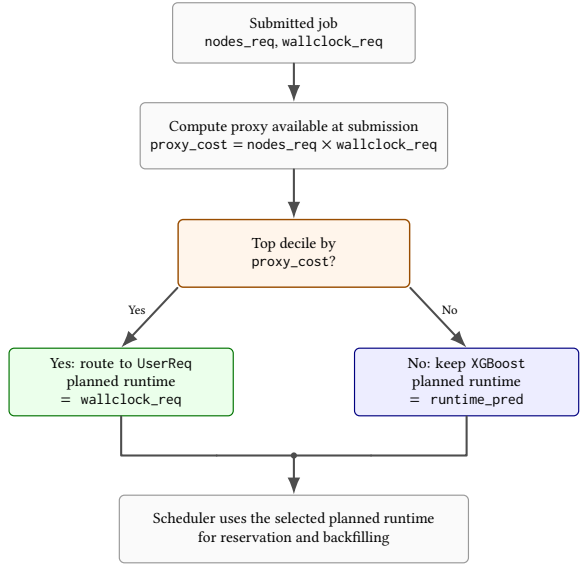
\begin{figure}[H]
\centering
\resizebox{0.9\columnwidth}{!}{\begin{tikzpicture}[font=\scriptsize, >=Latex, line cap=round, line join=round]

\tikzset{
  flowbox/.style={
    rounded corners=2pt,
    fill=black!2,
    draw=black!40,
    line width=0.45pt,
    align=center,
    inner sep=4pt
  },
  decision/.style={
    rounded corners=2pt,
    fill=orange!8,
    draw=orange!60!black,
    line width=0.55pt,
    align=center,
    inner sep=4pt
  },
  branchbox/.style={
    rounded corners=2pt,
    line width=0.45pt,
    align=center,
    inner sep=3pt,
    text width=2.7cm
  },
  link/.style={line width=0.8pt, draw=black!70},
  route/.style={->, line width=0.8pt, draw=black!70},
  lab/.style={font=\tiny, fill=white, inner sep=1pt}
}

\node[flowbox, minimum width=3.3cm, minimum height=0.82cm] (job) at (0,3.55)
  {Submitted job\\\texttt{nodes\_req}, \texttt{wallclock\_req}};

\node[flowbox, minimum width=4.15cm, minimum height=0.90cm] (proxy) at (0,2.15)
  {Compute proxy available at submission\\$\texttt{proxy\_cost}=\texttt{nodes\_req}\times\texttt{wallclock\_req}$};

\node[decision, minimum width=3.15cm, minimum height=0.92cm] (switch) at (0,0.55)
  {Top decile by\\\texttt{proxy\_cost}?};

\node[branchbox, fill=green!8, draw=green!45!black, minimum width=3.05cm, minimum height=0.92cm] (userreq) at (-2.35,-1.2)
  {Yes: route to \texttt{UserReq}\\planned runtime\\$=\texttt{wallclock\_req}$};

\node[branchbox, fill=blue!7, draw=blue!50!black, minimum width=3.05cm, minimum height=0.92cm] (xgb) at (2.35,-1.2)
  {No: keep \texttt{XGBoost}\\planned runtime\\$=\texttt{runtime\_pred}$};

\node[flowbox, minimum width=4.75cm, minimum height=0.92cm] (sched) at (0,-3.20)
  {Scheduler uses the selected planned runtime\\for reservation and backfilling};

\coordinate (branchleft) at (-1.05,-2.20);
\coordinate (branchright) at (1.05,-2.20);
\coordinate (merge) at (0,-2.20);
\coordinate (schedmerge) at ([yshift=0.22cm]sched.north);

\draw[route] (job.south) -- (proxy.north);
\draw[route] (proxy.south) -- (switch.north);
\draw[route] (switch.south west) -- node[lab, above left] {Yes} (userreq.north);
\draw[route] (switch.south east) -- node[lab, above right] {No} (xgb.north);
\draw[link] (userreq.south) |- (branchleft);
\draw[link] (xgb.south) |- (branchright);
\draw[link] (branchleft) -- (merge);
\draw[link] (branchright) -- (merge);
\fill[black!70] (merge) circle[radius=1.2pt];
\draw[route] (schedmerge) -- (sched.north);
\draw[link] (merge) -- (schedmerge);

\end{tikzpicture}}
\caption{The \texttt{Hybrid-p90} scheduling policy used in online simulation. Jobs above a p90 cutoff for each queue and year in \texttt{proxy\_cost} are routed to \texttt{UserReq}; the rest remain on \texttt{XGBoost}.}
\Description{A flow diagram for the Hybrid-p90 scheduling policy. A submitted job provides nodes\_req and wallclock\_req. The policy computes proxy\_cost as their product, compares it with a queue-year p90 threshold, routes jobs above that cutoff to UserReq and the rest to XGBoost, and then passes the selected planned runtime to the scheduler.}
\label{fig:hybrid-policy-flow}
\end{figure}

The hybrid policy applies \texttt{UserReq} only to jobs with high values of a proxy for resource usage available at submission. We use
\begin{equation}
\texttt{proxy\_cost} = \texttt{nodes\_req} \times \texttt{wallclock\_req}.
\end{equation}
Under \texttt{Hybrid-p90}, jobs above a p90 cutoff for each queue and year in \texttt{proxy\_cost} go to \texttt{UserReq}, and all others stay on \texttt{XGBoost}. This policy does not assume that \texttt{UserReq} should replace machine learning predictors everywhere. It tests whether routing the tail of jobs with high \texttt{proxy\_cost} to \texttt{UserReq} improves scheduling outcomes in online simulation. The p90 choice is intended as a simple, interpretable operating point motivated by the offline results rather than a claim of global optimality.
Figure~\ref{fig:hybrid-policy-flow} summarizes the routing rule.

\begin{figure*}[!t]
\centering
\includegraphics[width=\textwidth]{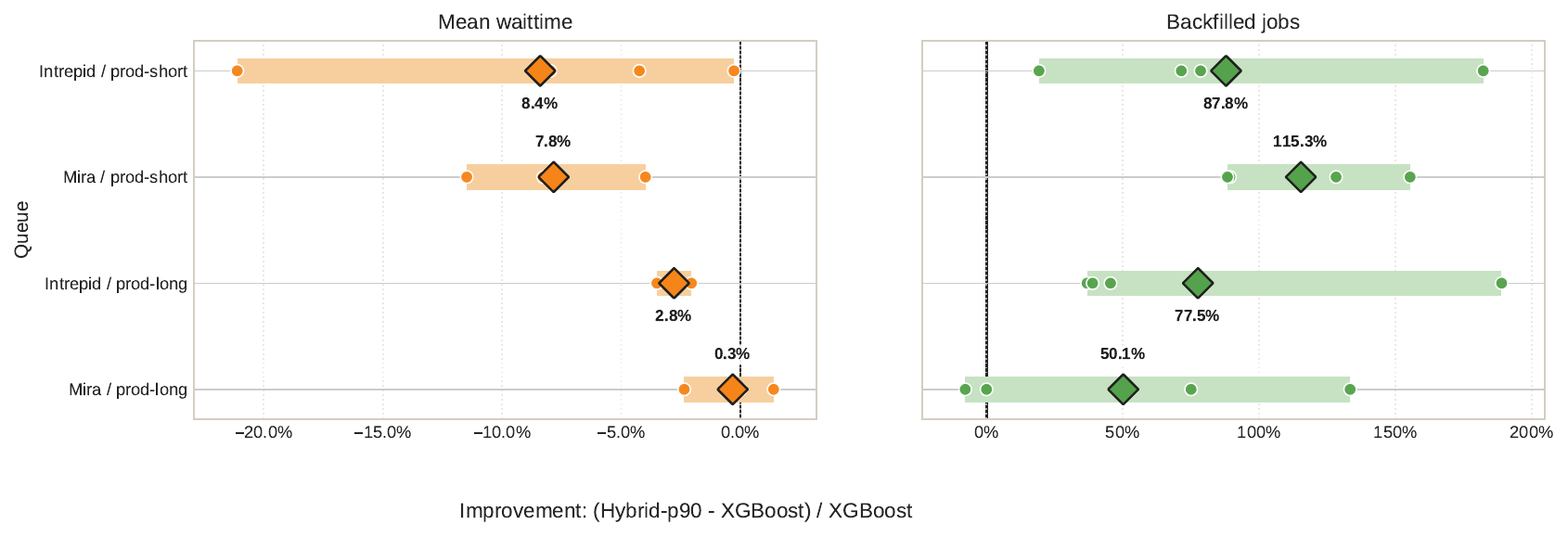}
\caption{Hybrid replay improves outcomes at the queue level across years. Each row is one queue. The pale background bar shows the span of yearly relative changes, the small points show individual years, and the diamond with its adjacent label marks the average for each queue under $(\texttt{Hybrid-p90} - \texttt{XGBoost}) / \texttt{XGBoost}$. Left: lower mean waittime is better. Right: more backfilled jobs are better.}
\Description{Two aligned dot-range panels summarize four queue-specific yearly online simulation cases. In each row, a pale horizontal background bar shows the minimum-to-maximum yearly relative change, small points show the individual years, and a larger diamond with a nearby percentage label shows the average across years. The mean waittime panel is favorable when values are below zero, and the backfilled jobs panel is favorable when values are above zero.}
\label{fig:hybrid-online-sim-cross-case}
\end{figure*}

\subsection{Online Simulation Setup}

To test the implications of the offline results for scheduling, we supplement the offline evaluation with online replay on \texttt{Mira} and \texttt{Intrepid}. We replay each queue and year partition independently, for example \texttt{Mira/prod-short/2016}. In this replay, the simulator preserves the original job chronology in timestamp order and changes only the planned runtime presented to the scheduler. Each job keeps its trace submit time and requested node count. The predictor under test supplies the planned runtime seen by the scheduler. For \texttt{XGBoost}, we use the same hyperparameter setting as in the original runtime prediction workflow of Menear et al.~\cite{menear2023MasteringHPC}. Reservation and backfilling decisions use that planned runtime, while node release uses the actual runtime from the trace.

The online simulation is intentionally simple. Each partition for a queue and year is simulated independently under \texttt{FCFS} with \texttt{EASY} backfilling. At each job submission or completion event, the scheduler first computes the earliest feasible reservation for the job at the head of the queue. It then scans later queued jobs in submit order and starts any job whose requested nodes and predicted runtime fit without delaying that reservation. Once started, a job occupies its requested nodes for its actual runtime. Those nodes are returned to the partition when the job completes. Partition capacity is inferred from the observed peak concurrent node usage in the trace.

To keep the comparison fair, the online simulation is restricted to jobs for which \texttt{XGBoost}, \texttt{Last2}, and \texttt{UserReq} all produce valid predictions after runtime filtering and prediction validity checks. The simulator reports mean wait time and the count of backfilled jobs.

For the online simulation comparison, we evaluate the hybrid routing policy defined above rather than treating the offline result as only a comparison at the model level. The online simulation summary reports the relative change in mean wait time and the count of backfilled jobs across four queues drawn from the long and short production queues on \texttt{Mira} and \texttt{Intrepid}. We focus on these four queues because short and long production queues are common and representative in HPC operations, and together they expose both short job and long job scheduling regimes on two different systems.

Across these four replay queues, \texttt{proxy\_cost} aligns well with the heavy tail region. At the same queue and year granularity used by the hybrid policy, the \texttt{proxy\_cost} top decile attains 80.4\%--91.2\% recall and 88.5\%--93.5\% recall weighted by cost with respect to the top decile of real cost. Table~\ref{tab:proxy-tail-detection} summarizes the values for each queue.

\begin{table}[t]
\centering
\caption{Precision, recall, and recall weighted by cost for each queue when the \texttt{proxy\_cost} top decile is used to identify the top decile of real cost.}
\label{tab:proxy-tail-detection}
\small
\begin{tabular}{llrrr}
\toprule
Dataset & Queue & Precision & Recall & \shortstack[r]{Cost-weighted\\Recall} \\
\midrule
Mira & prod-short & 72.9\% & 81.4\% & 89.8\% \\
Mira & prod-long & 57.5\% & 91.2\% & 93.5\% \\
Intrepid & prod-short & 61.1\% & 80.4\% & 88.5\% \\
Intrepid & prod-long & 69.5\% & 88.9\% & 92.4\% \\
\bottomrule
\end{tabular}

\end{table}

\subsection{Hybrid Replay Validates the Offline Signal}

We test whether the offline signal carries through to outcomes at the queue level on \texttt{Mira} and \texttt{Intrepid}. The hybrid rule routes only jobs with high \texttt{proxy\_cost} to \texttt{UserReq}. In the replay, \texttt{Hybrid-p90} sends the top decile of jobs by \texttt{proxy\_cost} within each partition for a queue and year to \texttt{UserReq} and leaves the rest on \texttt{XGBoost}. Figure~\ref{fig:hybrid-online-sim-cross-case} summarizes the comparison across cases in a single view. For each of four queues, the diamond and its adjacent label mark the average relative change across years under $(\texttt{Hybrid-p90} - \texttt{XGBoost}) / \texttt{XGBoost}$, the pale background bar shows the yearly span, and the smaller points show the individual cases for each queue and year. The yearly cases span 2015--2018 for \texttt{Mira} and 2010--2013 for \texttt{Intrepid}.

\texttt{Hybrid-p90} improves both metrics in all four cases. In every case, it reduces mean wait time relative to \texttt{XGBoost} and increases the number of backfilled jobs. The clearest gains in wait time appear in the short queues: \texttt{Intrepid/prod-short} shows an average relative wait reduction of 8.4\%, and \texttt{Mira/prod-short} shows a reduction of 7.8\%. \texttt{Intrepid/prod-long} still improves by 2.8\%, while \texttt{Mira/prod-long} is nearly neutral at 0.3\% lower average wait because one later year offsets earlier gains.

The backfilling side is even more uniformly favorable. The gains in short queues are especially large: 115.3\% in \texttt{Mira/prod-short} and 87.8\% in \texttt{Intrepid/prod-short}. The gains in long queues are also substantial. They reach 77.5\% in \texttt{Intrepid/prod-long} and 50.1\% in \texttt{Mira/prod-long}. These percentages should be interpreted directionally rather than as throughput multipliers, because the baseline number of backfilled jobs is small in some long queue years. Even with that caveat, the figure shows a clear and repeatable effect at the queue level: routing based on \texttt{proxy\_cost} tends to preserve more backfill opportunities while also shortening wait time. This pattern suggests that the short production queues leave more room for backfill gains in our replay, a result that is consistent with prior studies that characterize workloads~\cite{patel2020JobCharacteristics}.

The points for each year in Figure~\ref{fig:hybrid-online-sim-cross-case} show that the average diamonds are not driven by a single anomalous year. The short queues on both systems, together with \texttt{Intrepid/prod-long}, improve in every simulated year on both mean wait time and the count of backfilled jobs. The only mixed case is \texttt{Mira/prod-long}: 2016 and 2017 improve on both metrics, 2018 is slightly worse on both, and 2015 is effectively tied because both policies backfill the same small number of jobs. The dot range summary therefore preserves the yearly pattern directly instead of hiding instability behind a single average.

These results link the offline tail effect to scheduling outcomes. A simple policy at submission time that invokes \texttt{UserReq} only on jobs with high \texttt{proxy\_cost} improves both mean wait time and backfilling in all four queues. The simulation is intentionally limited: it uses partitions for each queue and year with \texttt{FCFS} scheduling and \texttt{EASY} backfilling rather than a production scheduler deployment. We therefore interpret the result as evidence of scheduling impact rather than full scheduler validation.

\section{Related Work}

We place this paper in the context of three adjacent areas: runtime prediction methods, studies of user provided walltime estimates and backfilling, and work that evaluates prediction quality through scheduling outcomes. This paper is closest to the third area because it asks how runtime predictors should be evaluated, rather than proposing a new predictor.

\subsection{HPC Runtime Prediction}

HPC job runtime prediction has a long history spanning models based on historical information, predictors based on logs, and more recent machine learning systems. Early work showed that historical behavior can support scheduling decisions~\cite{smith1998PredictingApplication}, and later studies used system logs or job attributes to predict completion time more directly~\cite{chen2013PredictingJob, wang2022PredictingJob}. Recent work has focused largely on improving the predictor itself through ensemble methods, deep learning, feature engineering, or domain specific and two step estimation strategies~\cite{chen2020RuntimePrediction, chen2023JobRuntime, cheon2023AREDAutomatabased, menear2023MasteringHPC, nunes2025TwoStepEstimation}. Among these, tree ensemble methods such as \texttt{XGBoost}~\cite{chen2016XGBoost} are especially common because they handle heterogeneous submission features and nonlinear interactions well, and they have become strong practical baselines in HPC runtime prediction studies~\cite{chen2020RuntimePrediction, menear2023MasteringHPC}. In this literature, performance is usually summarized with aggregate arithmetic mean metrics such as RMSE, MAE, or overall accuracy like scores, sometimes together with risk measures focused on underestimation~\cite{fan2017TradeOffPrediction}.

A related but distinct line of work studies predictive uncertainty through conditional quantiles and prediction intervals. Large scale quantile regression makes quantile estimation practical on very large datasets~\cite{yang2013QuantileRegressionLargeScale}, and conformalized quantile regression adds prediction intervals that are valid for finite samples~\cite{romano2019ConformalizedQuantile}. These methods target uncertainty quantification, not evaluation oriented toward schedulers under resource usage with heavy tails.

Our question is different from both predictor construction and interval estimation. We reevaluate existing prediction sources under different aggregation rules. Under resource usage with heavy tails, the choice of metric can reveal clearer differences among predictors and therefore support different conclusions about their quality for scheduling. The focus is therefore on predictor evaluation rather than predictor construction.

\subsection{User Provided Runtime Estimates and Backfilling}

User provided runtime estimates have been studied for decades because they affect reservation length, backfilling opportunities, and queue performance directly. Much of this literature emphasizes that user estimates are systematically overestimated or otherwise inaccurate in aggregate~\cite{baileylee2005AreUser, tsafrir2005ModelingUser}. However, the same line of work also shows that their operational value should not be dismissed: under \texttt{EASY} backfilling, conservative user estimates can preserve reservation safety and may even help utilization or queue behavior in some settings~\cite{feitelson1998UtilizationPredictability, mualem2001UtilizationPredictability, tsafrir2006DynamicsBackfilling, tsafrir2010UsingInaccurate}. Subsequent work studied adjusted user estimates, controls for underestimation, and scheduler mechanisms that replace or augment user estimates with predictions generated by the system~\cite{tang2010AnalyzingAdjusting, fan2017TradeOffPrediction, tsafrir2007BackfillingUsing, gaussier2015ImprovingBackfilling, chlumsky2022ImprovingAccuracy, cui2025ClusteringBased}. Runtime prediction has also been linked to queue time prediction and scheduler performance more broadly~\cite{smith1999UsingRunTime, menear2024TandemPredictions, vercellino2023MachineLearning, brown2024PredictingAccurate}.

Our results complement this literature. We do not claim that user estimates are the most useful prediction source across the whole workload. Instead, we show that they become especially valuable on the upper tail of resource usage and that this effect becomes visible only under evaluation focused on the tail.

\subsection{Evaluation Metrics and Scheduling Impact}

Prior work has argued that prediction quality should ultimately be judged by its effect on scheduling rather than by offline accuracy alone~\cite{downey1997PredictingQueue, smith1999UsingRunTime, sonmez2009TracebasedEvaluation}. Trace studies also show that large HPC workloads are heterogeneous and that aggregate averages can hide important structural differences across jobs, users, and resource classes~\cite{aaziz2018ModelingExpected, netti2021ConceptualFramework, patel2020JobCharacteristics}. At the same time, most runtime prediction papers still treat the evaluation metric as fixed and place the novelty in the predictor.

Relative to that evaluation oriented line of work, this paper combines three elements. First, it uses an overall metric, $A_{\mathrm{geo}}$, that weights error by resource usage alongside the metric based on averaging over jobs, $A_{\mathrm{mean}}$. Second, it uses equally sized deciles of resource usage to explain why GeoAccuracy reveals clearer differences among the methods. Third, it uses online replay to test whether those offline differences translate into scheduling outcomes. We are not aware of prior HPC runtime prediction work that centers this combination of evaluation focused on the tail, explanation at the decile level, and validation at the queue level, while explicitly linking the offline metric choice to the upper tail of the workload that matters most for scheduling.

\section{Conclusion}

This paper makes three contributions. First, it presents an empirical evaluation methodology for HPC job runtime prediction that focuses on the tail and combines GeoAccuracy weighted by resource usage with analyses at the decile and split levels. Second, it shows on three production HPC workloads that evaluation focused on the tail changes the offline conclusion. $A_{\mathrm{mean}}$ keeps the compared methods relatively close, whereas $A_{\mathrm{geo}}$ reveals clearer separation and makes \texttt{UserReq}'s strength in the resource dominant upper tail visible. In the top decile, \texttt{UserReq} achieves the highest GeoAccuracy and the lowest underestimation rate on all three datasets, and this advantage remains stable across rolling split days.

Third, it translates this offline signal into a simple hybrid scheduling policy and shows in online replay that it improves scheduling performance. This policy uses \texttt{proxy\_cost} at submission as the routing rule. It keeps \texttt{XGBoost} for most jobs and routes the top decile to \texttt{UserReq}. Across the short and long production queues on two HPC systems, it reduces mean wait time by up to 8\% and increases backfilled jobs by 50\%--115\%.

These results indicate that offline evaluation focused on the tail should inform the assessment of runtime predictors and scheduling policy design under HPC workloads with heavy tails. They also suggest that no single prediction source serves the entire workload equally well: machine learning predictors remain useful on the bulk of jobs, while the user provided walltime estimate at job submission carries valuable signal in the upper tail. Online schedulers should therefore combine multiple prediction signals through a hybrid policy rather than forcing one source to serve the whole workload.

\bibliographystyle{ACM-Reference-Format}
\bibliography{refs}


\begin{thebibliography}{35}


\ifx \showCODEN    \undefined \def \showCODEN     #1{\unskip}     \fi
\ifx \showISBNx    \undefined \def \showISBNx     #1{\unskip}     \fi
\ifx \showISBNxiii \undefined \def \showISBNxiii  #1{\unskip}     \fi
\ifx \showISSN     \undefined \def \showISSN      #1{\unskip}     \fi
\ifx \showLCCN     \undefined \def \showLCCN      #1{\unskip}     \fi
\ifx \shownote     \undefined \def \shownote      #1{#1}          \fi
\ifx \showarticletitle \undefined \def \showarticletitle #1{#1}   \fi
\ifx \showURL      \undefined \def \showURL       {\relax}        \fi
\providecommand\bibfield[2]{#2}
\providecommand\bibinfo[2]{#2}
\providecommand\natexlab[1]{#1}
\providecommand\showeprint[2][]{arXiv:#2}

\bibitem[Aaziz et~al\mbox{.}(2018)]%
        {aaziz2018ModelingExpected}
\bibfield{author}{\bibinfo{person}{Omar Aaziz}, \bibinfo{person}{Jonathan
  Cook}, {and} \bibinfo{person}{Mohammed Tanash}.}
  \bibinfo{year}{2018}\natexlab{}.
\newblock \showarticletitle{Modeling {{Expected Application Runtime}} for
  {{Characterizing}} and {{Assessing Job Performance}}}. In
  \bibinfo{booktitle}{\emph{2018 {{IEEE International Conference}} on {{Cluster
  Computing}} ({{CLUSTER}})}}. \bibinfo{publisher}{IEEE},
  \bibinfo{address}{Belfast}, \bibinfo{pages}{543--551}.
\newblock
\showISBNx{978-1-5386-8319-4}
\href{https://doi.org/10.1109/CLUSTER.2018.00070}{doi:\nolinkurl{10.1109/CLUSTER.2018.00070}}


\bibitem[Bailey~Lee et~al\mbox{.}(2005)]%
        {baileylee2005AreUser}
\bibfield{author}{\bibinfo{person}{Cynthia Bailey~Lee}, \bibinfo{person}{Yael
  Schwartzman}, \bibinfo{person}{Jennifer Hardy}, {and} \bibinfo{person}{Allan
  Snavely}.} \bibinfo{year}{2005}\natexlab{}.
\newblock \showarticletitle{Are {{User Runtime Estimates Inherently
  Inaccurate}}?}
\newblock In \bibinfo{booktitle}{\emph{Job {{Scheduling Strategies}} for
  {{Parallel Processing}}}}, \bibfield{editor}{\bibinfo{person}{David
  Hutchison}, \bibinfo{person}{Takeo Kanade}, \bibinfo{person}{Josef Kittler},
  \bibinfo{person}{Jon~M. Kleinberg}, \bibinfo{person}{Friedemann Mattern},
  \bibinfo{person}{John~C. Mitchell}, \bibinfo{person}{Moni Naor},
  \bibinfo{person}{Oscar Nierstrasz}, \bibinfo{person}{C.~Pandu~Rangan},
  \bibinfo{person}{Bernhard Steffen}, \bibinfo{person}{Madhu Sudan},
  \bibinfo{person}{Demetri Terzopoulos}, \bibinfo{person}{Dough Tygar},
  \bibinfo{person}{Moshe~Y. Vardi}, \bibinfo{person}{Gerhard Weikum},
  \bibinfo{person}{Dror~G. Feitelson}, \bibinfo{person}{Larry Rudolph}, {and}
  \bibinfo{person}{Uwe Schwiegelshohn}} (Eds.). Vol.~\bibinfo{volume}{3277}.
  \bibinfo{publisher}{JSSPP '05}, \bibinfo{address}{Berlin, Heidelberg},
  \bibinfo{pages}{253--263}.
\newblock
\showISBNx{978-3-540-25330-3 978-3-540-31795-1}
\href{https://doi.org/10.1007/11407522_14}{doi:\nolinkurl{10.1007/11407522_14}}


\bibitem[Brown et~al\mbox{.}(2024)]%
        {brown2024PredictingAccurate}
\bibfield{author}{\bibinfo{person}{Nick Brown}, \bibinfo{person}{Gordon Gibb},
  \bibinfo{person}{Evgenij Belikov}, {and} \bibinfo{person}{Rupert Nash}.}
  \bibinfo{year}{2024}\natexlab{}.
\newblock \showarticletitle{Predicting Accurate Batch Queue Wait Times on
  Production Supercomputers by Combining Machine Learning Techniques}.
\newblock \bibinfo{journal}{\emph{Concurrency and Computation: Practice and
  Experience}} \bibinfo{volume}{36}, \bibinfo{number}{15} (\bibinfo{date}{July}
  \bibinfo{year}{2024}), \bibinfo{pages}{e8112}.
\newblock
\showISSN{1532-0626, 1532-0634}
\href{https://doi.org/10.1002/cpe.8112}{doi:\nolinkurl{10.1002/cpe.8112}}


\bibitem[Chen(2023)]%
        {chen2023JobRuntime}
\bibfield{author}{\bibinfo{person}{Fengxian Chen}.}
  \bibinfo{year}{2023}\natexlab{}.
\newblock \showarticletitle{Job Runtime Prediction of {{HPC}} Cluster Based on
  {{PC-Transformer}}}.
\newblock \bibinfo{journal}{\emph{Journal of Supercomputing}}
  \bibinfo{volume}{79}, \bibinfo{number}{17} (\bibinfo{date}{Nov.}
  \bibinfo{year}{2023}), \bibinfo{pages}{20208--20234}.
\newblock
\showISSN{1573-0484}
\href{https://doi.org/10.1007/s11227-023-05470-2}{doi:\nolinkurl{10.1007/s11227-023-05470-2}}


\bibitem[Chen and Guestrin(2016)]%
        {chen2016XGBoost}
\bibfield{author}{\bibinfo{person}{Tianqi Chen} {and} \bibinfo{person}{Carlos
  Guestrin}.} \bibinfo{year}{2016}\natexlab{}.
\newblock \showarticletitle{{{XGBoost}}: A Scalable Tree Boosting System}. In
  \bibinfo{booktitle}{\emph{Proceedings of the 22nd {{ACM SIGKDD International
  Conference}} on {{Knowledge Discovery}} and {{Data Mining}}}}.
  \bibinfo{publisher}{ACM}, \bibinfo{address}{San Francisco, CA, USA},
  \bibinfo{pages}{785--794}.
\newblock
\href{https://doi.org/10.1145/2939672.2939785}{doi:\nolinkurl{10.1145/2939672.2939785}}


\bibitem[Chen et~al\mbox{.}(2013)]%
        {chen2013PredictingJob}
\bibfield{author}{\bibinfo{person}{Xin Chen}, \bibinfo{person}{Charng-Da Lu},
  {and} \bibinfo{person}{Karthik Pattabiraman}.}
  \bibinfo{year}{2013}\natexlab{}.
\newblock \showarticletitle{Predicting Job Completion Times Using System Logs
  in Supercomputing Clusters}. In \bibinfo{booktitle}{\emph{2013 43rd {{Annual
  IEEE}}/{{IFIP Conference}} on {{Dependable Systems}} and {{Networks
  Workshop}} ({{DSN-W}})}}. \bibinfo{publisher}{IEEE},
  \bibinfo{address}{Budapest, Hungary}, \bibinfo{pages}{1--8}.
\newblock
\showISBNx{978-1-4799-0181-4}
\href{https://doi.org/10.1109/DSNW.2013.6615513}{doi:\nolinkurl{10.1109/DSNW.2013.6615513}}


\bibitem[Chen et~al\mbox{.}(2020)]%
        {chen2020RuntimePrediction}
\bibfield{author}{\bibinfo{person}{Xiaomeng Chen}, \bibinfo{person}{Hui Zhang},
  \bibinfo{person}{Hanli Bai}, \bibinfo{person}{Chunming Yang},
  \bibinfo{person}{Xujian Zhao}, {and} \bibinfo{person}{Bo Li}.}
  \bibinfo{year}{2020}\natexlab{}.
\newblock \showarticletitle{Runtime Prediction of High-Performance Computing
  Jobs Based on Ensemble Learning}. In \bibinfo{booktitle}{\emph{Proceedings of
  the 2020 4th {{International Conference}} on {{High Performance
  Compilation}}, {{Computing}} and {{Communications}}}}.
  \bibinfo{publisher}{ACM}, \bibinfo{address}{Guangzhou China},
  \bibinfo{pages}{56--62}.
\newblock
\showISBNx{978-1-4503-7691-4}
\href{https://doi.org/10.1145/3407947.3407968}{doi:\nolinkurl{10.1145/3407947.3407968}}


\bibitem[Cheon et~al\mbox{.}(2023)]%
        {cheon2023AREDAutomatabased}
\bibfield{author}{\bibinfo{person}{Hyunjoon Cheon}, \bibinfo{person}{Jinseung
  Ryu}, \bibinfo{person}{Jaecheol Ryou}, \bibinfo{person}{Chan~Yeol Park},
  {and} \bibinfo{person}{Yo-Sub Han}.} \bibinfo{year}{2023}\natexlab{}.
\newblock \showarticletitle{{{ARED}}: Automata-Based Runtime Estimation for
  Distributed Systems Using Deep Learning}.
\newblock \bibinfo{journal}{\emph{Cluster Computing}} \bibinfo{volume}{26},
  \bibinfo{number}{5} (\bibinfo{date}{Oct.} \bibinfo{year}{2023}),
  \bibinfo{pages}{2629--2641}.
\newblock
\showISSN{1573-7543}
\href{https://doi.org/10.1007/s10586-021-03272-w}{doi:\nolinkurl{10.1007/s10586-021-03272-w}}


\bibitem[Chlumsk{\'y} and Klus{\'a}{\v c}ek(2022)]%
        {chlumsky2022ImprovingAccuracy}
\bibfield{author}{\bibinfo{person}{V{\'a}clav Chlumsk{\'y}} {and}
  \bibinfo{person}{Dalibor Klus{\'a}{\v c}ek}.}
  \bibinfo{year}{2022}\natexlab{}.
\newblock \showarticletitle{Improving {{Accuracy}} of {{Walltime Estimates}} in
  {{PBS Professional Using Soft Walltimes}}}.
\newblock In \bibinfo{booktitle}{\emph{Job {{Scheduling Strategies}} for
  {{Parallel Processing}}}}, \bibfield{editor}{\bibinfo{person}{Dalibor
  Klus{\'a}{\v c}ek}, \bibinfo{person}{Corbal{\'a}n Julita}, {and}
  \bibinfo{person}{Gonzalo~P. Rodrigo}} (Eds.). Vol.~\bibinfo{volume}{13592}.
  \bibinfo{publisher}{JSSPP '22}, \bibinfo{address}{Cham},
  \bibinfo{pages}{192--210}.
\newblock
\showISBNx{978-3-031-22697-7 978-3-031-22698-4}
\href{https://doi.org/10.1007/978-3-031-22698-4_10}{doi:\nolinkurl{10.1007/978-3-031-22698-4_10}}


\bibitem[Cui et~al\mbox{.}(2025)]%
        {cui2025ClusteringBased}
\bibfield{author}{\bibinfo{person}{Hang Cui}, \bibinfo{person}{Keichi
  Takahashi}, \bibinfo{person}{Yoichi Shimomura}, {and}
  \bibinfo{person}{Hiroyuki Takizawa}.} \bibinfo{year}{2025}\natexlab{}.
\newblock \showarticletitle{Clustering {{Based Job Runtime Prediction}} for
  {{Backfilling Using Classification}}}.
\newblock In \bibinfo{booktitle}{\emph{Job {{Scheduling Strategies}} for
  {{Parallel Processing}}}}, \bibfield{editor}{\bibinfo{person}{Dalibor
  Klus{\'a}{\v c}ek}, \bibinfo{person}{Julita Corbal{\'a}n}, {and}
  \bibinfo{person}{Gonzalo~P. Rodrigo}} (Eds.). Vol.~\bibinfo{volume}{14591}.
  \bibinfo{publisher}{JSSPP '24}, \bibinfo{address}{Cham},
  \bibinfo{pages}{40--59}.
\newblock
\showISBNx{978-3-031-74429-7 978-3-031-74430-3}
\href{https://doi.org/10.1007/978-3-031-74430-3_3}{doi:\nolinkurl{10.1007/978-3-031-74430-3_3}}


\bibitem[Downey(1997)]%
        {downey1997PredictingQueue}
\bibfield{author}{\bibinfo{person}{A.B. Downey}.}
  \bibinfo{year}{1997}\natexlab{}.
\newblock \showarticletitle{Predicting Queue Times on Space-Sharing Parallel
  Computers}. In \bibinfo{booktitle}{\emph{Proceedings 11th {{International
  Parallel Processing Symposium}}}}. \bibinfo{publisher}{IEEE Comput. Soc.
  Press}, \bibinfo{address}{Genva, Switzerland}, \bibinfo{pages}{209--218}.
\newblock
\showISBNx{978-0-8186-7793-9}
\href{https://doi.org/10.1109/IPPS.1997.580894}{doi:\nolinkurl{10.1109/IPPS.1997.580894}}


\bibitem[Fan et~al\mbox{.}(2017)]%
        {fan2017TradeOffPrediction}
\bibfield{author}{\bibinfo{person}{Yuping Fan}, \bibinfo{person}{Paul Rich},
  \bibinfo{person}{William~E. Allcock}, \bibinfo{person}{Michael~E. Papka},
  {and} \bibinfo{person}{Zhiling Lan}.} \bibinfo{year}{2017}\natexlab{}.
\newblock \showarticletitle{Trade-{{Off Between Prediction Accuracy}} and
  {{Underestimation Rate}} in {{Job Runtime Estimates}}}. In
  \bibinfo{booktitle}{\emph{2017 {{IEEE International Conference}} on {{Cluster
  Computing}} ({{CLUSTER}})}}. \bibinfo{publisher}{IEEE},
  \bibinfo{address}{Honolulu, HI, USA}, \bibinfo{pages}{530--540}.
\newblock
\showISBNx{978-1-5386-2326-8}
\href{https://doi.org/10.1109/CLUSTER.2017.11}{doi:\nolinkurl{10.1109/CLUSTER.2017.11}}


\bibitem[Feitelson and Weil(1998)]%
        {feitelson1998UtilizationPredictability}
\bibfield{author}{\bibinfo{person}{D.G. Feitelson} {and} \bibinfo{person}{A.M.
  Weil}.} \bibinfo{year}{1998}\natexlab{}.
\newblock \showarticletitle{Utilization and Predictability in Scheduling the
  {{IBM SP2}} with Backfilling}. In \bibinfo{booktitle}{\emph{Proceedings of
  the {{First Merged International Parallel Processing Symposium}} and
  {{Symposium}} on {{Parallel}} and {{Distributed Processing}}}}.
  \bibinfo{publisher}{IEEE Comput. Soc}, \bibinfo{address}{Orlando, FL, USA},
  \bibinfo{pages}{542--546}.
\newblock
\showISBNx{978-0-8186-8404-3}
\href{https://doi.org/10.1109/IPPS.1998.669970}{doi:\nolinkurl{10.1109/IPPS.1998.669970}}


\bibitem[Gaussier et~al\mbox{.}(2015)]%
        {gaussier2015ImprovingBackfilling}
\bibfield{author}{\bibinfo{person}{Eric Gaussier}, \bibinfo{person}{David
  Glesser}, \bibinfo{person}{Valentin Reis}, {and} \bibinfo{person}{Denis
  Trystram}.} \bibinfo{year}{2015}\natexlab{}.
\newblock \showarticletitle{Improving Backfilling by Using Machine Learning to
  Predict Running Times}. In \bibinfo{booktitle}{\emph{Proceedings of the
  {{International Conference}} for {{High Performance Computing}},
  {{Networking}}, {{Storage}} and {{Analysis}}}}. \bibinfo{publisher}{ACM},
  \bibinfo{address}{Austin Texas}, \bibinfo{pages}{1--10}.
\newblock
\showISBNx{978-1-4503-3723-6}
\href{https://doi.org/10.1145/2807591.2807646}{doi:\nolinkurl{10.1145/2807591.2807646}}


\bibitem[Li et~al\mbox{.}(2023)]%
        {li2023AnalyzingResource}
\bibfield{author}{\bibinfo{person}{Jie Li}, \bibinfo{person}{George
  Michelogiannakis}, \bibinfo{person}{Brandon Cook}, \bibinfo{person}{Dulanya
  Cooray}, {and} \bibinfo{person}{Yong Chen}.} \bibinfo{year}{2023}\natexlab{}.
\newblock \bibinfo{title}{Analyzing {{Resource Utilization}} in an {{HPC
  System}}: {{A Case Study}} of {{NERSC Perlmutter}}}.
\newblock
\showeprint[arxiv]{2301.05145}~[cs]
\href{https://doi.org/10.48550/arXiv.2301.05145}{doi:\nolinkurl{10.48550/arXiv.2301.05145}}


\bibitem[Lifka(1995)]%
        {lifka1995ANLIBM}
\bibfield{author}{\bibinfo{person}{David~A. Lifka}.}
  \bibinfo{year}{1995}\natexlab{}.
\newblock \showarticletitle{The {{ANL}}/{{IBM SP}} Scheduling System}.
\newblock In \bibinfo{booktitle}{\emph{Job {{Scheduling Strategies}} for
  {{Parallel Processing}}}}, \bibfield{editor}{\bibinfo{person}{Gerhard Goos},
  \bibinfo{person}{Juris Hartmanis}, \bibinfo{person}{Jan Leeuwen},
  \bibinfo{person}{Dror~G. Feitelson}, {and} \bibinfo{person}{Larry Rudolph}}
  (Eds.). Vol.~\bibinfo{volume}{949}. \bibinfo{publisher}{JSSPP '95},
  \bibinfo{address}{Berlin, Heidelberg}, \bibinfo{pages}{295--303}.
\newblock
\showISBNx{978-3-540-60153-1 978-3-540-49459-1}
\href{https://doi.org/10.1007/3-540-60153-8_35}{doi:\nolinkurl{10.1007/3-540-60153-8_35}}


\bibitem[Menear et~al\mbox{.}(2024)]%
        {menear2024TandemPredictions}
\bibfield{author}{\bibinfo{person}{Kevin Menear}, \bibinfo{person}{Kadidia
  Konate}, \bibinfo{person}{Kristi Potter}, {and} \bibinfo{person}{Dmitry
  Duplyakin}.} \bibinfo{year}{2024}\natexlab{}.
\newblock \showarticletitle{Tandem {{Predictions}} for {{HPC}} Jobs}. In
  \bibinfo{booktitle}{\emph{Practice and {{Experience}} in {{Advanced Research
  Computing}} 2024: {{Human Powered Computing}}}}. \bibinfo{publisher}{ACM},
  \bibinfo{address}{Providence RI USA}, \bibinfo{pages}{1--9}.
\newblock
\showISBNx{979-8-4007-0419-2}
\href{https://doi.org/10.1145/3626203.3670547}{doi:\nolinkurl{10.1145/3626203.3670547}}


\bibitem[Menear et~al\mbox{.}(2023)]%
        {menear2023MasteringHPC}
\bibfield{author}{\bibinfo{person}{Kevin Menear}, \bibinfo{person}{Ambarish
  Nag}, \bibinfo{person}{Jordan {Perr-Sauer}}, \bibinfo{person}{Monte Lunacek},
  \bibinfo{person}{Kristi Potter}, {and} \bibinfo{person}{Dmitry Duplyakin}.}
  \bibinfo{year}{2023}\natexlab{}.
\newblock \showarticletitle{Mastering {{HPC Runtime Prediction}}: {{From
  Observing Patterns}} to a {{Methodological Approach}}}. In
  \bibinfo{booktitle}{\emph{Practice and {{Experience}} in {{Advanced Research
  Computing}}}}. \bibinfo{publisher}{ACM}, \bibinfo{address}{Portland OR USA},
  \bibinfo{pages}{75--85}.
\newblock
\showISBNx{978-1-4503-9985-2}
\href{https://doi.org/10.1145/3569951.3593598}{doi:\nolinkurl{10.1145/3569951.3593598}}


\bibitem[Mu'alem and Feitelson(2001)]%
        {mualem2001UtilizationPredictability}
\bibfield{author}{\bibinfo{person}{A.W. Mu'alem} {and} \bibinfo{person}{D.G.
  Feitelson}.} \bibinfo{year}{2001}\natexlab{}.
\newblock \showarticletitle{Utilization, Predictability, Workloads, and User
  Runtime Estimates in Scheduling the {{IBM SP2}} with Backfilling}.
\newblock \bibinfo{journal}{\emph{TPDS '01}} \bibinfo{volume}{12},
  \bibinfo{number}{6} (\bibinfo{date}{June} \bibinfo{year}{2001}),
  \bibinfo{pages}{529--543}.
\newblock
\showISSN{10459219}
\href{https://doi.org/10.1109/71.932708}{doi:\nolinkurl{10.1109/71.932708}}


\bibitem[Netti et~al\mbox{.}(2021)]%
        {netti2021ConceptualFramework}
\bibfield{author}{\bibinfo{person}{Alessio Netti}, \bibinfo{person}{Woong
  Shin}, \bibinfo{person}{Michael Ott}, \bibinfo{person}{Torsten Wilde}, {and}
  \bibinfo{person}{Natalie Bates}.} \bibinfo{year}{2021}\natexlab{}.
\newblock \showarticletitle{A {{Conceptual Framework}} for {{HPC Operational
  Data Analytics}}}. In \bibinfo{booktitle}{\emph{2021 {{IEEE International
  Conference}} on {{Cluster Computing}} ({{CLUSTER}})}}.
  \bibinfo{publisher}{IEEE}, \bibinfo{address}{Portland, OR, USA},
  \bibinfo{pages}{596--603}.
\newblock
\showISBNx{978-1-7281-9666-4}
\href{https://doi.org/10.1109/Cluster48925.2021.00086}{doi:\nolinkurl{10.1109/Cluster48925.2021.00086}}


\bibitem[Nunes et~al\mbox{.}(2025)]%
        {nunes2025TwoStepEstimation}
\bibfield{author}{\bibinfo{person}{Alan~L. Nunes}, \bibinfo{person}{Bernardo
  Gallo}, \bibinfo{person}{Bruno Lopes}, \bibinfo{person}{Felipe~A. Portella},
  \bibinfo{person}{Jos{\'e} Viterbo}, \bibinfo{person}{L{\'u}cia M.~A.
  Drummond}, \bibinfo{person}{Luciano Andrade}, \bibinfo{person}{Miguel {de
  Lima}}, \bibinfo{person}{Paulo J.~B. Estrela}, {and}
  \bibinfo{person}{Renzo~Q. Malini}.} \bibinfo{year}{2025}\natexlab{}.
\newblock \showarticletitle{Two-{{Step Estimation Strategy}} for {{Predicting
  Petroleum Reservoir Simulation Jobs Runtime}} on an {{HPC Cluster}}}.
\newblock \bibinfo{journal}{\emph{CPE}} \bibinfo{volume}{37},
  \bibinfo{number}{4-5} (\bibinfo{year}{2025}), \bibinfo{pages}{e70026}.
\newblock
\showISSN{1532-0634}
\href{https://doi.org/10.1002/cpe.70026}{doi:\nolinkurl{10.1002/cpe.70026}}


\bibitem[Patel et~al\mbox{.}(2020)]%
        {patel2020JobCharacteristics}
\bibfield{author}{\bibinfo{person}{Tirthak Patel}, \bibinfo{person}{Zhengchun
  Liu}, \bibinfo{person}{Raj Kettimuthu}, \bibinfo{person}{Paul Rich},
  \bibinfo{person}{William Allcock}, {and} \bibinfo{person}{Devesh Tiwari}.}
  \bibinfo{year}{2020}\natexlab{}.
\newblock \showarticletitle{Job {{Characteristics}} on {{Large-Scale Systems}}:
  {{Long-Term Analysis}}, {{Quantification}}, and {{Implications}}}. In
  \bibinfo{booktitle}{\emph{{{SC20}}: {{International Conference}} for {{High
  Performance Computing}}, {{Networking}}, {{Storage}} and {{Analysis}}}}.
  \bibinfo{publisher}{IEEE}, \bibinfo{address}{Atlanta, GA, USA},
  \bibinfo{pages}{1--17}.
\newblock
\showISBNx{978-1-7281-9998-6}
\href{https://doi.org/10.1109/SC41405.2020.00088}{doi:\nolinkurl{10.1109/SC41405.2020.00088}}


\bibitem[Romano et~al\mbox{.}(2019)]%
        {romano2019ConformalizedQuantile}
\bibfield{author}{\bibinfo{person}{Yaniv Romano}, \bibinfo{person}{Evan
  Patterson}, {and} \bibinfo{person}{Emmanuel~J. Candes}.}
  \bibinfo{year}{2019}\natexlab{}.
\newblock \showarticletitle{Conformalized Quantile Regression}. In
  \bibinfo{booktitle}{\emph{Advances in Neural Information Processing Systems
  32}}. \bibinfo{publisher}{Curran Associates, Inc.}, \bibinfo{address}{Red
  Hook, NY, USA}, \bibinfo{pages}{3538--3548}.
\newblock


\bibitem[Smith et~al\mbox{.}(1998)]%
        {smith1998PredictingApplication}
\bibfield{author}{\bibinfo{person}{Warren Smith}, \bibinfo{person}{Ian Foster},
  {and} \bibinfo{person}{Valerie Taylor}.} \bibinfo{year}{1998}\natexlab{}.
\newblock \showarticletitle{Predicting Application Run Times Using Historical
  Information}.
\newblock In \bibinfo{booktitle}{\emph{Job {{Scheduling Strategies}} for
  {{Parallel Processing}}}}, \bibfield{editor}{\bibinfo{person}{Gerhard Goos},
  \bibinfo{person}{Juris Hartmanis}, \bibinfo{person}{Jan Van~Leeuwen},
  \bibinfo{person}{Dror~G. Feitelson}, {and} \bibinfo{person}{Larry Rudolph}}
  (Eds.). Vol.~\bibinfo{volume}{1459}. \bibinfo{publisher}{JSSPP '98},
  \bibinfo{address}{Berlin, Heidelberg}, \bibinfo{pages}{122--142}.
\newblock
\showISBNx{978-3-540-64825-3 978-3-540-68536-4}
\href{https://doi.org/10.1007/BFb0053984}{doi:\nolinkurl{10.1007/BFb0053984}}


\bibitem[Smith et~al\mbox{.}(1999)]%
        {smith1999UsingRunTime}
\bibfield{author}{\bibinfo{person}{Warren Smith}, \bibinfo{person}{Valerie
  Taylor}, {and} \bibinfo{person}{Ian Foster}.}
  \bibinfo{year}{1999}\natexlab{}.
\newblock \showarticletitle{Using {{Run-Time Predictions}} to {{Estimate Queue
  Wait Times}} and {{Improve Scheduler Performance}}}.
\newblock In \bibinfo{booktitle}{\emph{Job {{Scheduling Strategies}} for
  {{Parallel Processing}}}}, \bibfield{editor}{\bibinfo{person}{Gerhard Goos},
  \bibinfo{person}{Juris Hartmanis}, \bibinfo{person}{Jan Van~Leeuwen},
  \bibinfo{person}{Dror~G. Feitelson}, {and} \bibinfo{person}{Larry Rudolph}}
  (Eds.). Vol.~\bibinfo{volume}{1659}. \bibinfo{publisher}{JSSPP '99},
  \bibinfo{address}{Berlin, Heidelberg}, \bibinfo{pages}{202--219}.
\newblock
\showISBNx{978-3-540-66676-9 978-3-540-47954-3}
\href{https://doi.org/10.1007/3-540-47954-6_11}{doi:\nolinkurl{10.1007/3-540-47954-6_11}}


\bibitem[Sonmez et~al\mbox{.}(2009)]%
        {sonmez2009TracebasedEvaluation}
\bibfield{author}{\bibinfo{person}{Ozan Sonmez}, \bibinfo{person}{Nezih
  Yigitbasi}, \bibinfo{person}{Alexandru Iosup}, {and} \bibinfo{person}{Dick
  Epema}.} \bibinfo{year}{2009}\natexlab{}.
\newblock \showarticletitle{Trace-Based Evaluation of Job Runtime and Queue
  Wait Time Predictions in Grids}. In \bibinfo{booktitle}{\emph{Proceedings of
  the 18th {{ACM}} International Symposium on {{High}} Performance Distributed
  Computing}}. \bibinfo{publisher}{ACM}, \bibinfo{address}{Garching Germany},
  \bibinfo{pages}{111--120}.
\newblock
\showISBNx{978-1-60558-587-1}
\href{https://doi.org/10.1145/1551609.1551632}{doi:\nolinkurl{10.1145/1551609.1551632}}


\bibitem[Tang et~al\mbox{.}(2010)]%
        {tang2010AnalyzingAdjusting}
\bibfield{author}{\bibinfo{person}{Wei Tang}, \bibinfo{person}{Narayan Desai},
  \bibinfo{person}{Daniel Buettner}, {and} \bibinfo{person}{Zhiling Lan}.}
  \bibinfo{year}{2010}\natexlab{}.
\newblock \showarticletitle{Analyzing and Adjusting User Runtime Estimates to
  Improve Job Scheduling on the {{Blue Gene}}/{{P}}}. In
  \bibinfo{booktitle}{\emph{2010 {{IEEE International Symposium}} on
  {{Parallel}} \& {{Distributed Processing}} ({{IPDPS}})}}.
  \bibinfo{publisher}{IEEE}, \bibinfo{address}{Atlanta, GA, USA},
  \bibinfo{pages}{1--11}.
\newblock
\showISBNx{978-1-4244-6442-5}
\href{https://doi.org/10.1109/IPDPS.2010.5470474}{doi:\nolinkurl{10.1109/IPDPS.2010.5470474}}


\bibitem[Tsafrir(2010)]%
        {tsafrir2010UsingInaccurate}
\bibfield{author}{\bibinfo{person}{Dan Tsafrir}.}
  \bibinfo{year}{2010}\natexlab{}.
\newblock \showarticletitle{Using {{Inaccurate Estimates Accurately}}}.
\newblock In \bibinfo{booktitle}{\emph{Job {{Scheduling Strategies}} for
  {{Parallel Processing}}}}, \bibfield{editor}{\bibinfo{person}{Eitan
  Frachtenberg} {and} \bibinfo{person}{Uwe Schwiegelshohn}} (Eds.).
  Vol.~\bibinfo{volume}{6253}. \bibinfo{publisher}{JSSPP '10},
  \bibinfo{address}{Berlin, Heidelberg}, \bibinfo{pages}{208--221}.
\newblock
\showISBNx{978-3-642-16504-7 978-3-642-16505-4}
\href{https://doi.org/10.1007/978-3-642-16505-4_12}{doi:\nolinkurl{10.1007/978-3-642-16505-4_12}}


\bibitem[Tsafrir et~al\mbox{.}(2005)]%
        {tsafrir2005ModelingUser}
\bibfield{author}{\bibinfo{person}{Dan Tsafrir}, \bibinfo{person}{Yoav Etsion},
  {and} \bibinfo{person}{Dror~G. Feitelson}.} \bibinfo{year}{2005}\natexlab{}.
\newblock \showarticletitle{Modeling {{User Runtime Estimates}}}.
\newblock In \bibinfo{booktitle}{\emph{Job {{Scheduling Strategies}} for
  {{Parallel Processing}}}}, \bibfield{editor}{\bibinfo{person}{David
  Hutchison}, \bibinfo{person}{Takeo Kanade}, \bibinfo{person}{Josef Kittler},
  \bibinfo{person}{Jon~M. Kleinberg}, \bibinfo{person}{Friedemann Mattern},
  \bibinfo{person}{John~C. Mitchell}, \bibinfo{person}{Moni Naor},
  \bibinfo{person}{Oscar Nierstrasz}, \bibinfo{person}{C.~Pandu~Rangan},
  \bibinfo{person}{Bernhard Steffen}, \bibinfo{person}{Madhu Sudan},
  \bibinfo{person}{Demetri Terzopoulos}, \bibinfo{person}{Dough Tygar},
  \bibinfo{person}{Moshe~Y. Vardi}, \bibinfo{person}{Gerhard Weikum},
  \bibinfo{person}{Dror Feitelson}, \bibinfo{person}{Eitan Frachtenberg},
  \bibinfo{person}{Larry Rudolph}, {and} \bibinfo{person}{Uwe Schwiegelshohn}}
  (Eds.). Vol.~\bibinfo{volume}{3834}. \bibinfo{publisher}{JSSPP '05},
  \bibinfo{address}{Berlin, Heidelberg}, \bibinfo{pages}{1--35}.
\newblock
\showISBNx{978-3-540-31024-2 978-3-540-31617-6}
\href{https://doi.org/10.1007/11605300_1}{doi:\nolinkurl{10.1007/11605300_1}}


\bibitem[Tsafrir et~al\mbox{.}(2007)]%
        {tsafrir2007BackfillingUsing}
\bibfield{author}{\bibinfo{person}{Dan Tsafrir}, \bibinfo{person}{Yoav Etsion},
  {and} \bibinfo{person}{Dror~G. Feitelson}.} \bibinfo{year}{2007}\natexlab{}.
\newblock \showarticletitle{Backfilling {{Using System-Generated Predictions
  Rather}} than {{User Runtime Estimates}}}.
\newblock \bibinfo{journal}{\emph{TPDS '07}} \bibinfo{volume}{18},
  \bibinfo{number}{6} (\bibinfo{date}{June} \bibinfo{year}{2007}),
  \bibinfo{pages}{789--803}.
\newblock
\showISSN{1045-9219}
\href{https://doi.org/10.1109/TPDS.2007.70606}{doi:\nolinkurl{10.1109/TPDS.2007.70606}}


\bibitem[Tsafrir and Feitelson(2006)]%
        {tsafrir2006DynamicsBackfilling}
\bibfield{author}{\bibinfo{person}{Dan Tsafrir} {and} \bibinfo{person}{Dror~G.
  Feitelson}.} \bibinfo{year}{2006}\natexlab{}.
\newblock \showarticletitle{The {{Dynamics}} of {{Backfilling}}: {{Solving}}
  the {{Mystery}} of {{Why Increased Inaccuracy May Help}}}. In
  \bibinfo{booktitle}{\emph{2006 {{IEEE International Symposium}} on {{Workload
  Characterization}}}}. \bibinfo{publisher}{IEEE}, \bibinfo{address}{San Jose,
  CA}, \bibinfo{pages}{131--141}.
\newblock
\showISBNx{978-1-4244-0508-4}
\href{https://doi.org/10.1109/IISWC.2006.302737}{doi:\nolinkurl{10.1109/IISWC.2006.302737}}


\bibitem[Vercellino et~al\mbox{.}(2023)]%
        {vercellino2023MachineLearning}
\bibfield{author}{\bibinfo{person}{Chiara Vercellino}, \bibinfo{person}{Alberto
  Scionti}, \bibinfo{person}{Giuseppe Varavallo}, \bibinfo{person}{Paolo
  Viviani}, \bibinfo{person}{Giacomo Vitali}, {and} \bibinfo{person}{Olivier
  Terzo}.} \bibinfo{year}{2023}\natexlab{}.
\newblock \showarticletitle{A {{Machine Learning Approach}} for an {{HPC Use
  Case}}: The {{Jobs Queuing Time Prediction}}}.
\newblock \bibinfo{journal}{\emph{Future Generation Computer Systems}}
  \bibinfo{volume}{143} (\bibinfo{date}{June} \bibinfo{year}{2023}),
  \bibinfo{pages}{215--230}.
\newblock
\showISSN{0167739X}
\href{https://doi.org/10.1016/j.future.2023.01.020}{doi:\nolinkurl{10.1016/j.future.2023.01.020}}


\bibitem[Wang et~al\mbox{.}(2021)]%
        {wang2021UserlevelWorkload}
\bibfield{author}{\bibinfo{person}{Qiqi Wang}, \bibinfo{person}{Yu Shen}, {and}
  \bibinfo{person}{Jing Li}.} \bibinfo{year}{2021}\natexlab{}.
\newblock \showarticletitle{User-Level {{Workload Analysis}} for
  {{Supercomputers}}}. In \bibinfo{booktitle}{\emph{2021 {{The}} 4th
  {{International Conference}} on {{Software Engineering}} and {{Information
  Management}}}}. \bibinfo{publisher}{ACM}, \bibinfo{address}{Yokohama Japan},
  \bibinfo{pages}{68--73}.
\newblock
\showISBNx{978-1-4503-8895-5}
\href{https://doi.org/10.1145/3451471.3451483}{doi:\nolinkurl{10.1145/3451471.3451483}}


\bibitem[Wang et~al\mbox{.}(2022)]%
        {wang2022PredictingJob}
\bibfield{author}{\bibinfo{person}{Qiqi Wang}, \bibinfo{person}{Hongjie Zhang},
  \bibinfo{person}{Jing Li}, \bibinfo{person}{Yu Shen}, {and}
  \bibinfo{person}{Xiaohui Liu}.} \bibinfo{year}{2022}\natexlab{}.
\newblock \showarticletitle{Predicting Job Finish Time Based on Parameter
  Features and Running Logs in Supercomputing System}.
\newblock \bibinfo{journal}{\emph{The Journal of Supercomputing}}
  \bibinfo{volume}{78}, \bibinfo{number}{17} (\bibinfo{date}{Nov.}
  \bibinfo{year}{2022}), \bibinfo{pages}{18551--18577}.
\newblock
\showISSN{0920-8542, 1573-0484}
\href{https://doi.org/10.1007/s11227-022-04582-5}{doi:\nolinkurl{10.1007/s11227-022-04582-5}}


\bibitem[Yang et~al\mbox{.}(2013)]%
        {yang2013QuantileRegressionLargeScale}
\bibfield{author}{\bibinfo{person}{Jiyan Yang}, \bibinfo{person}{Xiangrui
  Meng}, {and} \bibinfo{person}{Michael~W. Mahoney}.}
  \bibinfo{year}{2013}\natexlab{}.
\newblock \showarticletitle{Quantile Regression for Large-Scale Applications}.
  In \bibinfo{booktitle}{\emph{Proceedings of the 30th International Conference
  on Machine Learning}}, Vol.~\bibinfo{volume}{28}. \bibinfo{publisher}{PMLR},
  \bibinfo{address}{Atlanta, Georgia, USA}, \bibinfo{pages}{881--887}.
\newblock


\end{thebibliography}

\end{document}